\documentclass{emulateapj}

\begin{document}
 
\title{Imprints of Environment on Cluster and Field Late-Type Galaxies at
$z\sim 1$}
 
\author{N.~L. Homeier\altaffilmark{1}
M. Postman\altaffilmark{2},
F. Menanteau\altaffilmark{1},
J.~P. Blakeslee\altaffilmark{1}
S. Mei\altaffilmark{1},
R. Demarco\altaffilmark{1},
H.~C. Ford\altaffilmark{1},
G.~D. Illingworth\altaffilmark{3},
A. Zirm\altaffilmark{4}
}
\altaffiltext{1}{Department of Physics and Astronomy, Johns Hopkins
University, 3400 North Charles Street, Baltimore, MD 21218.}
 \altaffiltext{2}{STScI, 3700 San Martin Drive, Baltimore, MD 21218.}
\altaffiltext{3}{UCO/Lick Observatory, University of California, Santa
Cruz, CA 95064.}
\altaffiltext{4}{Leiden Observatory, Postbus 9513, 2300 RA Leiden,
Netherlands.}

\date{Accepted by the Astronomical Journal}
 
\begin{abstract}
We present a comparison of late-type galaxies (Sa and later) in 
intermediate redshift clusters and the field using images from
the Advanced Camera for Surveys (ACS) aboard the {\it Hubble Space Telescope}
(HST). Cluster and field galaxies are selected by matching photometric
and spectroscopic catalogs of four cluster fields:
CL0152-1357, CL1056-0337 (MS1054), CL1604+4304, and CL1604+4321. 
Concentration, asymmetry, and clumpiness parameters
are calculated for each galaxy in blue (F606W or F625W) and red 
(F775W or F814W) filters. Galaxy half-light radii, disk scale lengths,
color gradients, and overall color are compared. We find marginally 
significant differences 
in the asymmetry distributions of spiral and irregular galaxies in the
X-ray luminous and X-ray faint clusters. The massive clusters contain 
fewer galaxies with large asymmetries. The physical sizes of the cluster 
and field populations are similar; no significant differences are found 
in half-light radii or disk scale lengths. The most significant difference is
in rest-frame $U-B$ color. Late-type cluster galaxies are
significantly redder, $\sim 0.3$ magnitudes at rest-frame $U-B$, 
than their field counterparts. Moreover, 
the intermediate-redshift cluster galaxies tend to have blue inward 
color gradients, in contrast to the field galaxies, but similar to late-type
galaxies in low redshift clusters. These blue inward color gradients are likely
to be the result of enhanced nuclear star formation rates relative to the
outer disk. Based on the significant rest-frame color difference, we 
conclude that late-type cluster members at $z\sim0.9$
are not a pristine infalling field population; some difference in past 
and/or current star 
formation history is already present. This points to high redshift 
``groups'', or filaments
with densities similar to present-day groups, as the sites where the first
major effects of environment are imprinted. 

\end{abstract}
 

\maketitle
 
\section{Introduction}

Understanding the physical processes that shape present day galaxies 
is one of modern astronomy's 
fundamental goals. Galaxy morphology is clearly related to environment
\citep{Dressler80,PG84,WG91}. Few gas-rich galaxies and relatedly, few 
galaxies with spiral morphologies are found in the cores of rich 
clusters. To what extent the environment drives galaxy 
evolution (nurture) or is simply a roadmap for the distribution of 
galaxy halo masses (nature) is being currently refined.

By observing clusters at high redshift we can
directly observe dynamically young structures. There is substantial
evidence that overall cluster galaxy populations evolve with redshift.
Fractionally more blue galaxies are found in clusters at 
higher redshifts; this trend is the Butcher-Oemler (B-O) effect \citep{BO84}. 
However, cluster cores do not show a star-forming galaxy B-O 
effect \citep{Nakataetal05}, and the B-O effect depends sensitively on
cluster radius \citep{Ellingsonetal01,Nakataetal05, Wakeetal05}. This implies 
that the Butcher-Oemler galaxies are infalling field galaxies, and indeed,
some blue cluster galaxies have been identified as such \citep{Tranetal05}. 
We test whether the late-type population in intermediate
redshift clusters is an infalling field population 
by comparing the properties of cluster and field galaxies 
at $z\sim0.9$,
currently the highest redshift where a substantial number of 
clusters are known.

%\subsection{Galaxy Evolution and Environment}

At low redshift there is a color-density relation (e.g. Goto et~al. 2004)
similar to the well-known morphology-density
relation \citep{Dressler80}, and widespread evidence that 
star formation rates are lower in cluster and group galaxies than in the field
\citep{Baloghetal97,Baloghetal98,Hashimotoetal98,Cetal01,
Lewisetal02,Martinezetal02}. The morphology-density,
color-density, and star formation rate (SFR)-density relations may be
caused by physical processes such as 
galaxy-galaxy interactions and tidal stripping \citep{Mooreetal98},
starvation \citep{Larsonetal80,Bekkietal02}, and/or ram pressure stripping 
\citep{GG72,Quilisetal00,SS01}. Alternatively, 
morphology, color, and SFR could be determined by the galaxy's
halo mass, and this correlates strongly with galaxy density. 
Fortunately, large surveys have been able to address this issue.
Environment is important in determining the evolution of all but the 
most massive galaxies (with M~$>3\times10^{10}$~M$_{\odot}$)
\citep{Kauffmannetal04,Tanakaetal04}, but we have yet to determine the 
physical mechanisms which are responsible for these observed 
environmental effects.  

The study of galaxy properties and environment from the SDSS by 
\citet{Kauffmannetal04} has shown
that the galaxy property that is most sensitive to environment is
star formation history. 
Star formation in galaxies less massive than $3\times10^10$~M$_{\odot}$
and in regions of enhanced local ($< 1$~Mpc) galaxy density
appears to decline gradually over a long, 1-3~Gyr, timescale. 
This is indicated by the lack of variation 
in the relationships between indicators of current and recent star 
formation in all environments.
This has important implications for the dominant transformation process(es).
Ram pressure stripping is thought to operate on 
short, $<1$~Gyr timescales, while other mechanisms, such as starvation or
galaxy harassment, are thought to shut off star formation on longer timescales.
However, as a caveat, recent simulations show that a truncated gas disk 
can persist for a few Gyr \citep{RH05}.

Another way to determine which physical processes drive 
galaxy evolution is to identify the physical characteristics of the
regions in which galaxy properties change.
Several groups have reported a ``break'' in the otherwise smooth distribution
of the fraction of galaxies with H$\alpha$ emission with local galaxy 
density \citep{Gomezetal03,Baloghetal04,Tanakaetal04}, with fewer 
H$\alpha$-emitters in higher density regions. However, among the 
population with significant star formation, no correlation between 
H$\alpha$ equivalent 
width and density was found. The interpretation is that the timescale
for the transition from star-forming to non-star-forming is rapid, 
$< 1$~Gyr. This is in conflict with the results of 
\citet{Kauffmannetal04} that star formation is gradually extinguished. 

Extending these studies to higher redshifts allows us to observe how
these trends evolve and trace how they are established.
We now know that the morphology-density relation exists at
intermediate redshifts \citep{Dressleretal97,Smithetal04,Postmanetal05},
and evolves smoothly with redshift \citep{Smithetal04,Postmanetal05}.
However, the elliptical fraction 
as a function of density does not evolve significantly \citep{Postmanetal05},
and the change in the morphology-density relation between 
$z\sim1$ and the present-day universe is in the relative fractions
of S0 and spiral+irregular galaxies.
The trend of reduced SFR for galaxies in clusters is also established
at these redshifts \citep{Ellingsonetal01,Postmanetal01}, although
the evolution of total cluster SFRs with redshift is not yet clear 
\citep{Finnetal04,Kodamaetal04,Homeieretal05,Finnetal05}. 

Detailed morphological measurements of cluster galaxies at $z\sim 1$
have recently become possible with the installation of the Advanced Camera
for Surveys (ACS; Ford et~al. 2003) on the HST. The ACS intermediate 
redshift cluster survey probes 7 clusters in the redshift range 
$0.83 \leq z \leq 1.27$. Previous papers in this series
have discussed the evolution
of the cluster color-magnitude relation at $z=1.24$ \citep{Blakesleeetal04}, 
the fundamental plane \citep{Holdenetal05a},
the size-surface brightness relation for early-type cluster galaxies 
\citep{Holdenetal05b}, the star-forming cluster galaxy population 
\citep{Homeieretal05}, the cluster 
galaxy luminosity function \citep{Gotoetal05}, and 
the morphology-density relation \citep{Postmanetal05}. In this paper
we explore morphological similarities and differences between spiral and 
irregular galaxies at intermediate redshifts in galaxy clusters and in the 
field. We compare quantitative morphological measurements, physical sizes, and 
colors of late-type galaxies in two X-ray luminous clusters 
at $z=0.84$, two X-ray faint clusters at $z=0.9$, and field galaxies 
at comparable redshifts. We aim to uncover whether these late-type 
cluster members have properties that make them
distinct from late-type field galaxies, or if they are
indistinguishable, in which case they are consistent with a pristine 
infalling field population. If they are already distinct from the field
population, it supports the scenario where environmental changes impact
galaxies on long, $> 1$~Gyr timescales. 

\section{Observations and Reductions}

CL0152-1357 (CL0152), CL1056-0337 (MS1054), CL1604+4304, and CL1604+4321 
were observed with the ACS 
Wide Field Channel as part of a guaranteed time observation 
program (proposal 9290). CL1604+4304 and CL1604+4321 were observed
with a single pointing in the $V_{606}$ and $I_{814}$ filters 
for two orbits each. For CL0152-1357 and MS1054, the observations 
were taken in a $2\times2$ (four pointing) mosaic pattern, with 2 
orbits of integration in the $r_{625}$ (CL0152), 1 orbit of $V_{606}$ 
(MS1054), and 2 orbits each of $i_{775}$ and 
$z_{850}$ filters. 
The cluster cores were imaged for 
a total of 32 orbits in each filter because of the $1\arcmin$ of
overlap between the pointings. The data were processed with the 
$Apsis$ pipeline \citep{Blakesleeetal03}. 
Our photometry is calibrated to the AB
magnitude system using zeropoints in Sirianni et~al. (2005). Object detection 
and photometry was performed by SExtractor \citep{BA96}
incorporated within the $Apsis$ pipeline. A more detailed description can be
found in \citet{Benitezetal04}. We use isophotal magnitudes, MAG\_ISO, 
when quoting colors as they provide a more accurate measure of galaxy
color, and MAG\_AUTO when quoting broad-band magnitudes, as this is the 
best estimate of a galaxy's total magnitude \citep{Benitezetal04}.

\subsection{Sample Selection}

\begin{figure}
\begin{center}
\includegraphics[width=9cm]{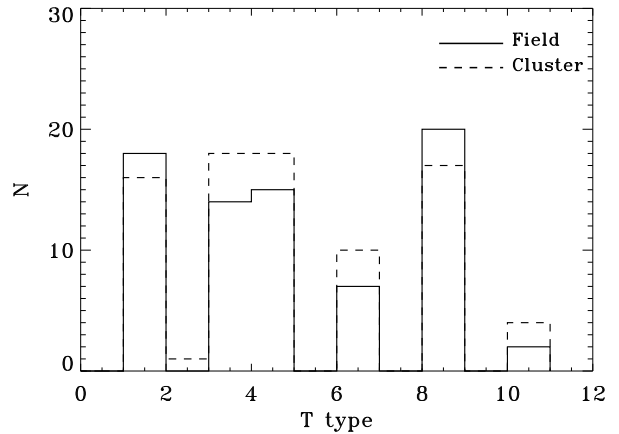}
\caption{The visually classified T-type distributions for the combined
cluster (dashed) and field (solid) samples. There is
no significant difference in the T-type distributions. The median T-type 
for the cluster sample is 4, and also for the field sample, 4. This 
illustrates that there is no significant difference in visual morphology
between the field and cluster samples.
  \label{fig:ttypes}}
\end{center}
\end{figure}

We list the properties of our sample galaxies in 
Tables~\ref{tab:4304cl}-\ref{tab:0152f},
and show color cutouts in Figures~\ref{fig:4304cl}-\ref{fig:0152f}.
\textbf{Figures 10-17 will appear in the online manuscript, and are not
in the astro-ph version.}
Spectroscopic redshift catalogs were used to select cluster and field samples
for CL0152-1357 \citep{Demarcoetal05}, MS1054 \citep{Tranetal99},
CL1604+4304 and CL1604+4321 \citep{Postmanetal98,Lubinetal98,Postmanetal01}.
Extensive visual morphology catalogs were created by MP 
\citep{Postmanetal05}. Morphologies were 
determined visually 
on the T-type system \citep{deVacetal91}. All galaxies in
the field with $i_{775}$ or $I_{814} \le 24$ magnitude were classified. 
Approximately 10\% of the galaxies were also classified by three
independent classifiers to estimate the classification errors. 
Majority agreement was achieved for 75\% of objects with 
$i_{775}\le 23.5$. There was no significant offset between
the mean classification from the independent classifiers. More information
can be found in \citet{Postmanetal05}.

A T-type $-5 \leq T \leq -3$ corresponds to elliptical galaxies, 
$-2 \leq$~T~$\leq 0$
corresponds to lenticular (S0) galaxies, and 
T~$\geq1$ to Sa and later-type galaxies.
We selected field galaxies with visual morphological type Sa or later, 
T-type $\geq 1$, and with redshifts between $0.55 \leq z \leq 1.1$,
and excluding the cluster redshift. Histograms of the T-type distributions
of cluster and field galaxies are shown in Figure~\ref{fig:ttypes}.
There are no significant differences in the
distribution of visually assigned T-types between any of the four 
cluster galaxy samples and their respective field samples. The median
T-type of each sample is $3-4$, except for the CL1604+4304 cluster and 
field samples, 
which have medians of 6. The median T-type for the combined 
cluster sample is 4, and for the combined field sample, also 4. 

The field galaxy
redshifts were obtained with the same masks as the cluster galaxy redshifts.
The advantage of using a field galaxy sample which is taken from the same
images as the cluster galaxies is that it minimizes the chance of
systematic errors in population properties.
The redshift completeness functions for MS1054, CL1604+4304, and CL1604+4321
depend only on $R$ magnitude. There is a color term in the redshift
completeness function for CL0152-1357. The mask selection was based on
photometric redshift, and galaxies bluer than the
red cluster sequence are less likely to have been observed 
\citep{Demarcoetal05,Homeieretal05}.
About $1/3$ of the spectroscopically
confirmed late-type population is within $3\sigma$ of the red cluster
sequence \citet{Homeieretal05}. Also, in the CL0152-1357 redshift
catalog, there is a galaxy group at $z\sim0.64$ \citep{Demarcoetal05}. 
We excluded 9 galaxies
with redshifts $0.62 < z < 0.65$ which were likely to be associated with
this group. Our final combined field sample contains 71 galaxies.

\subsection{CAS Parameters}

We measured the Concentration (C), Asymmetry (A), and Clumpiness (S) 
parameters \citep{Abrahametal94,Bershadyetal00,Conselice03}
for all cluster and field galaxies in our sample. Concentration is
related to galaxy mass, asymmetry is related 
to interactions and mergers, and clumpiness is related 
to the current star formation rate \citep{Conselice03}. The degree
of concentration, asymmetry, or clumpiness increases as the value
of the corresponding parameter increases.
A public version of the CAS code is available at http://acs.pha.jhu.edu/~felipe/PYCA and described further in \citep{Menanteauetal05}.

\subsubsection{C - Concentration}

The concentration definition we use is from \citet{Abrahametal94}.
Concentration is defined as the sum of the galaxy flux within 
$r_{0.3}=0.3\times r_{total}$ divided by the total flux. We fit an
ellipse to the Sextractor segmentation map; this is the aperture used
for the total flux. The segmentation map includes all pixels assigned
to the galaxy that are $1.5\sigma$ above the background.
The inner radius, $r_{0.3}$, is this aperture with the semi-major and
semi-minor axes multipled by 0.3. 
%As an equation, 
%C$=\frac{2!pi \int_0.3R^0 I(R)dR}{2!pi \int_R^0 I(R)dR} 

\subsubsection{A - Asymmetry}

Qualitatively, one calculates asymmetry by subtracting an 180-degree rotated
image from the original image, summing the residuals, and including a 
correction for the background. We smooth each galaxy image with a gaussian 
kernel with a width of 1 pixel. This smoothed image is rotated by 180 degrees
and subtracted from the smoothed, non-rotated image. The asymmetry forumla we
use is, $A = \frac{1}{2}\frac{\Sigma(|I - I_{rot}|) - B_{corr}}{I_{t}}$, 
where $I$ is the smoothed image, $I_{rot}$ is this image rotated by 
180 degrees, and $B_{corr}$ is a correction factor for the asymmetry
signal of the background. $B_{corr}$ = $\sqrt{2}\times Area \times SKYRMS$, 
and $I_{t}$ is the sum of the flux in the smoothed image. The asymmetry 
calculation uses only the pixels included in the Sextractor segmentation map.

\subsubsection{S - Clumpiness}

The clumpiness parameter is a measure of the high frequency residuals
in a galaxy image. In our clumpiness calculation we subtract the 
Sextractor-created background image from 
the galaxy image. This background-subtracted image is 
smoothed with a gaussian kernel with FWHM equal to 
5\% of the total radius (square root of the product of the 
semi-major and semi-minor axes of the Sextractor Kron aperture). The 
sum of the pixel values of this background-subtracted, smoothed image 
is divided by the sum of the pixel values unsmoothed background-subtracted 
image. 
Only pixels in the Sextractor segmentation map are used.
A formula for the clumpiness parameter can be expressed as, 
S$=10\times(\frac{\Sigma(I-I_{back})_{smooth}}{\Sigma(I-I_{back})}\times mask)$,
where the mask assures that negative pixels are set to zero before 
summing and that the central $3\times3$ pixels are excluded. The 
central region must be excluded to avoid obtaining anomalously high clumpiness 
values for galaxies which simply have larger central light 
concentrations (bulges). What we are interested in is the 
high frequency light variations from the outer regions of the galaxy, which
are related to current star formation rate.
We chose $3\times3$ pixels because it excludes the most problematic region
for the majority of our galaxies. This value is then multiplied by 10.

\section{Results}

\subsection{Colors}

\begin{figure*}
\begin{center}
\includegraphics[width=18cm]{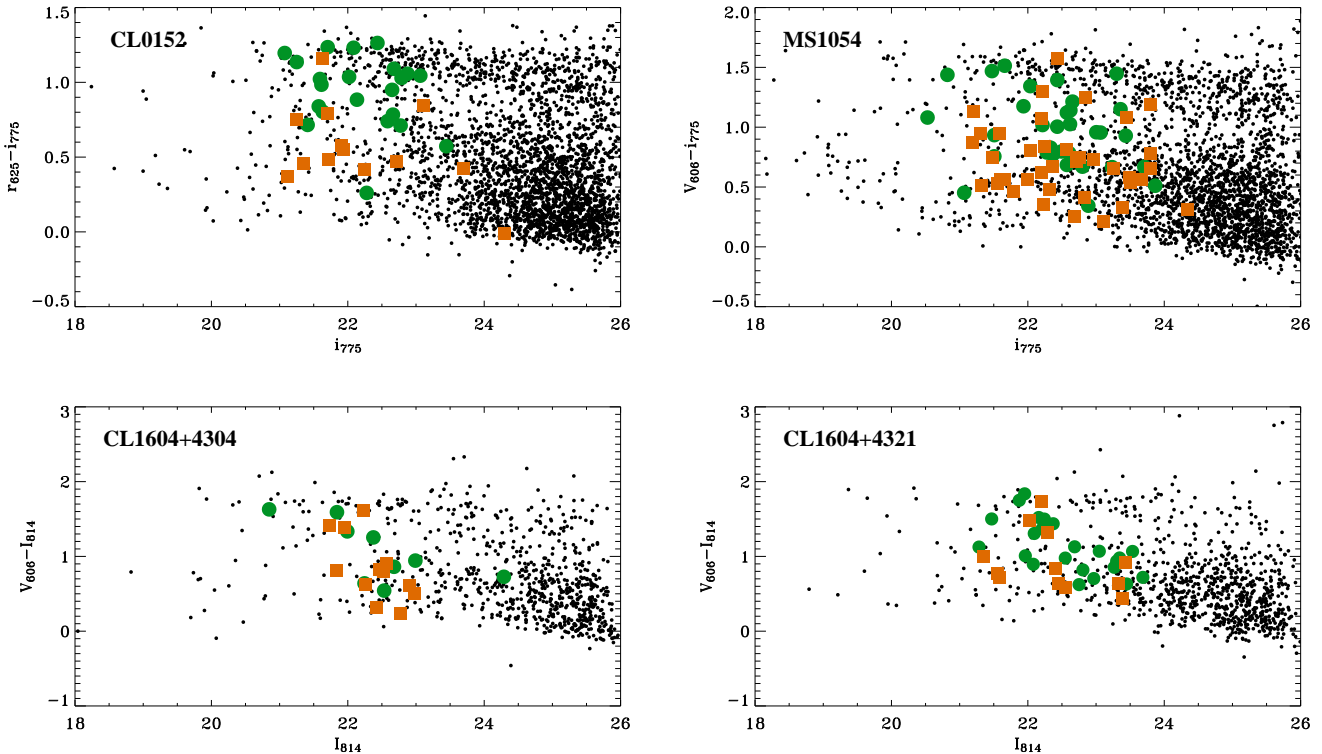}
\caption{Color magnitude diagrams for the cluster fields
with the spectroscopically confirmed late-type cluster members
as filled dots and the late-type field galaxies as filled squares.  
This illustrates the colors of the sample galaxies relative to all objects
in the field. Note that some of the late-type galaxies are in the 
red cluster sequence.
  \label{fig:cmds}}
\end{center}
\end{figure*}

\begin{figure*}
\begin{center}
\includegraphics[width=18cm]{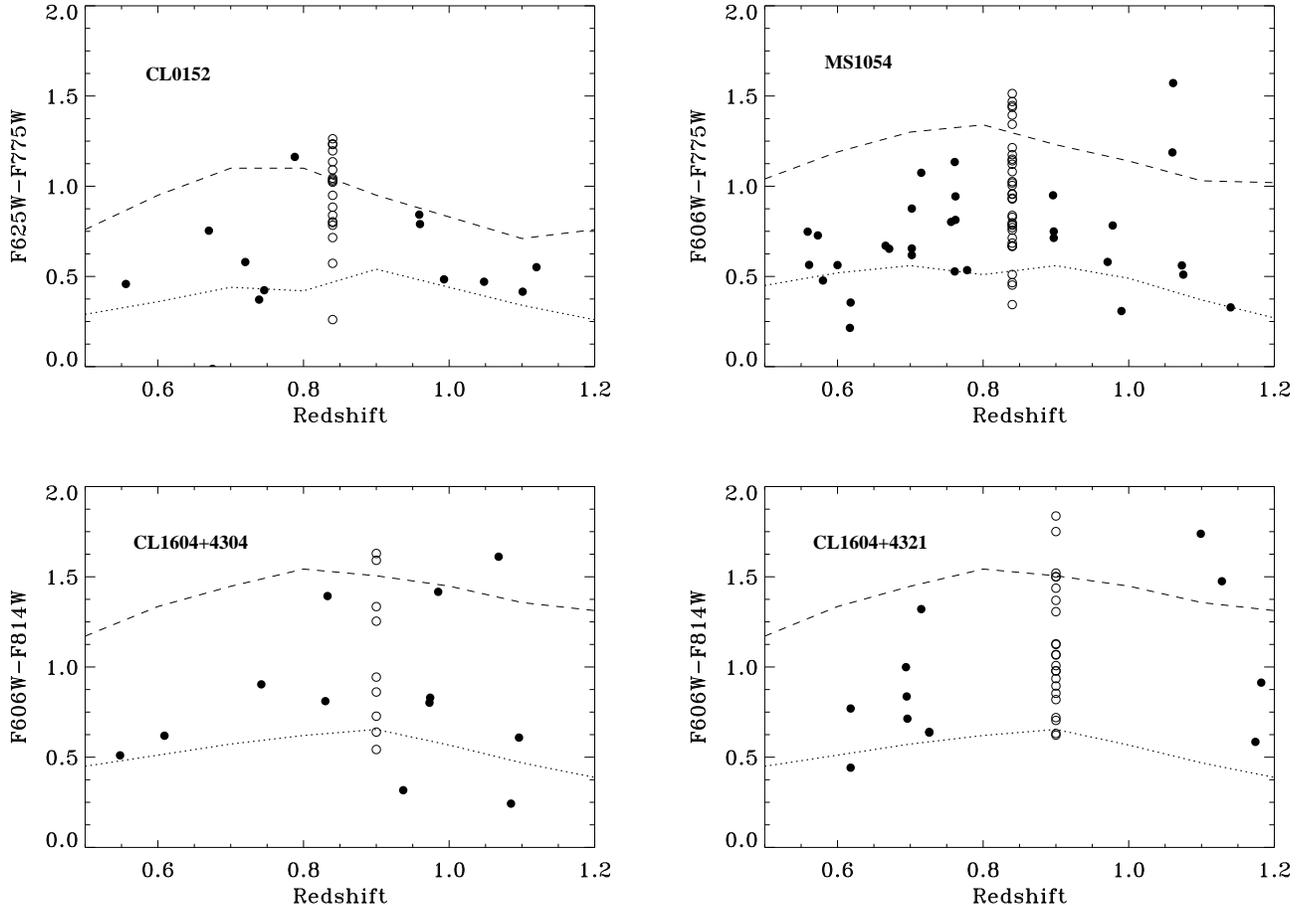}
\caption{Observed ACS color versus redshift for the field (solid dots) and
cluster (open dots) galaxies in our sample. Overplotted are the Kinney-Calzetti
Sb (dotted) and Starb1 (dashed) templates.
  \label{colorz}}
\end{center}
\end{figure*}

\begin{figure}
\begin{center}
\includegraphics[width=9cm]{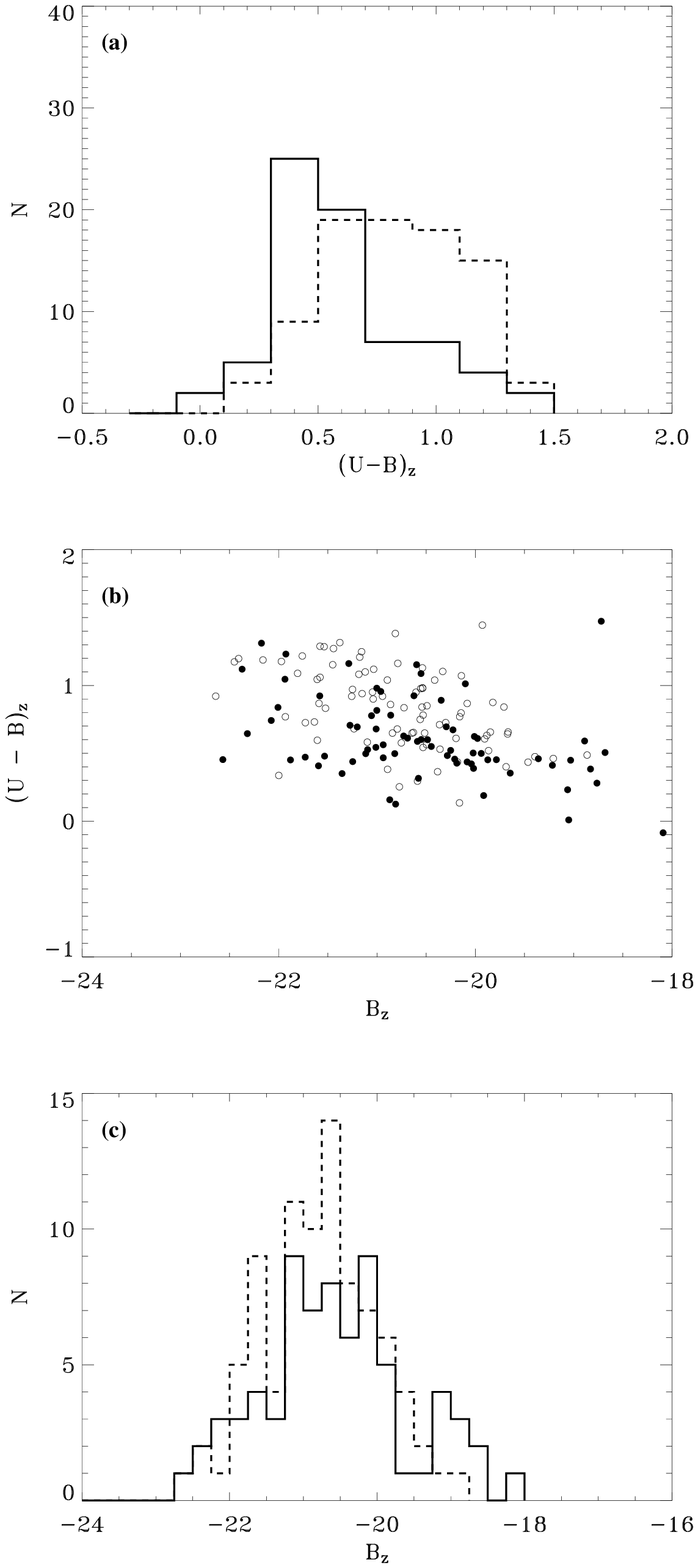}
\caption{Rest-frame U-B vs. B color-magnitude diagram of field (solid line, 
filled dots) and cluster (dashed line, open dots) galaxies. There is a 
significant
difference in rest-frame $U-B$ color between the cluster and field samples.
Although there is also a difference in the range of absolute $B$ magnitude 
(panel (c)), in panel (b) we can see that at a given $B$ magnitude the cluster
galaxies are significantly redder. See text for a discussion of the 
high significance of this color difference.   
  \label{fig:cmd}}
\end{center}
\end{figure}

Rest-frame color is a basic parameter of a galaxy and it reflects 
the integrated
star formation history. To compare the colors of the cluster and field
galaxies we convert observed magnitudes to rest-frame magnitudes and
then colors.
We transformed the observed ACS $r_{625}-i_{775}$ (CL0152), $V_{606}-i_{775}$
(MS1054) and $V_{606}-I_{814}$ (CL1604+4304,+4321) to rest-frame
$U-B$ (AB magnitudes) colors as follows. Using the IRAF task CALCPHOT, 
we redshifted the Kinney-Calzetti templates (S0, Sa, Sb, Starb1, and Starb2;
Kinney et~al. 1996) to the redshift of the galaxy
and calculated the observed ACS magnitude using the appropriate
filter transmission curves. Also using CALPHOT, we calculated $U$ and
$B$ at $z=0$. We then performed a linear fit of the 
observed ACS color (x-axis) with the rest-frame $U$ or $B$ magnitude minus
the observed ACS magnitude (y-axis). In other words, because the observed
ACS filters are close to rest-frame $U$ and $B$ filters, we can robustly
calculate the difference between the observed ACS magnitude and the 
rest-frame $U$ or $B$ magnitude. These ``corrections'' depend somewhat on the
spectral slope, which is probed by the observed color.
The ``corrections'' are of the order of 0.5 magnitudes for both filters.
In Figure~\ref{colorz} we show the observed colors and redshifts
of our sample galaxies
and how the observed colors for the Sb (dashed line) and Starb1 (solid line) 
templates vary with redshift. This illustrates that the chosen templates
reasonably cover the range of expected spectral slopes of our sample galaxies.

\begin{deluxetable}{lccc}
\tablecolumns{4}
\tablewidth{0pc}
\tablecaption{Color Transformations
\label{transtab}}
\tabletypesize{\scriptsize}
\tablehead{ 
\colhead{Cluster} & \colhead{Restframe Band} & \colhead{m} & \colhead{b}}
\startdata
CL0152      & U & $-0.14$ & 0.56 \\
CL0152      & B & $-0.20$ & 0.61 \\
MS1054      & U & $-0.30$ & 0.55 \\
MS1054      & B & $-0.17$ & 0.61 \\
CL1604+4304,21 & U & $-0.26$ & 0.61 \\
CL1604+4304,21 & B & $-0.16$ & 0.75 \\
\enddata
\end{deluxetable}

The terms in the linear transformation equations are listed in 
Table~\ref{transtab}. The transformations are of the form

\begin{equation}
U = m \times c + b + mag_{blue} 
\end{equation}

\begin{equation}
B = m \times c + b + mag_{red} 
\end{equation}

where c is the observed color and mag$_{blue}$ and mag$_{red}$  
are the observed magnitudes. For the cluster galaxies we 
used two transformation equations (for $U$ and $B$)
at redshifts 0.84 (CL0152 and MS1054) and 0.9 (CL1604+4304, +4321).
Colors are calculated using the Sextractor isophotal magnitudes, but 
mag$_{blue}$ and mag$_{red}$ are Sextractor MAG\_AUTO magnitudes.

In Figure~\ref{fig:cmd} we show a $(U-B)_{z}$ histogram, a rest-frame
color-magnitude diagram, and a $B_{z}$ histogram for
cluster (dashed line, open dots) and field (solid line, filled dots) 
galaxies. The field late-type sample is significantly
bluer than the cluster late-type sample. Although they also have a slightly
different distribution of absolute $B$ magnitude, at a given absolute
magnitude the cluster galaxies are redder, as can be seen in the middle panel
of Figure~\ref{fig:cmd}. There is no significant difference in color 
distribution between the late-type galaxies in the X-ray luminous and 
X-ray faint clusters.

The mean rest-frame $U-B$ colors of the cluster and field samples are 
0.83 and 0.59 magnitudes, respectively. 
This difference is highly significant, the K-S test definitely
ruling out the null hypothesis that the two samples are drawn from the 
same parent population, even if we restrict the samples to galaxies
with $B_{z} \leq -20$ (confidence level greater than 99.99\%).

%with a probability of belonging to the same parent population of less than 
%0.0001\% \textbf{with a K-S test}. If we restrict the samples to galaxies 
%with $B_{z} \leq -20$,
%the mean colors of the cluster and field samples are 0.86 and 0.66, still
%with only a 0.006\% probability of being drawn from the same parent 
%population, or in other words, a confidence limit of $>99.99$\%.

However, as mentioned in \S~2.1, the redshift completeness for the CL0152 
cluster sample is not uniform with magnitude; redder galaxies were 
more likely to have been observed. This should affect both the 
cluster and field sample in the CL0152 field, but 
if we are conservative and exclude the CL0152 cluster galaxies from the 
combined cluster sample, we
still find a highly significant difference between the cluster and field
samples. In this case, the mean rest-frame $U-B$ color of the cluster 
sample is 0.77 magnitudes, compared to 0.59 magnitudes for the field. 
This gives confidence limit of 99.98\%. Including the magnitude cut 
of $B_{z} \leq -20$, we have a confidence limit of 99.7\%.

 This difference persists when the late-type
galaxies in the red cluster sequence are excluded.
If we include only those galaxies with $(U-B)_{z} \leq 1$, which removes
the red cluster sequence members, this color difference is still highly
significant
(99.8\%). In this case the mean rest-frame $U-B$ colors are 0.68 and 0.53
magnitudes for the cluster and field samples, respectively. 
Our robust result is that there is a significant offset in 
rest-frame $U-B$ color between
the cluster and field late-type galaxies, in the sense that the cluster 
galaxies are redder. 
We found no correlation between rest-frame $U-B$ color and
cluster radius or local galaxy density.

\subsection{Cluster-Field CAS Comparison}

\begin{figure*}
\begin{center}
\includegraphics[width=18cm]{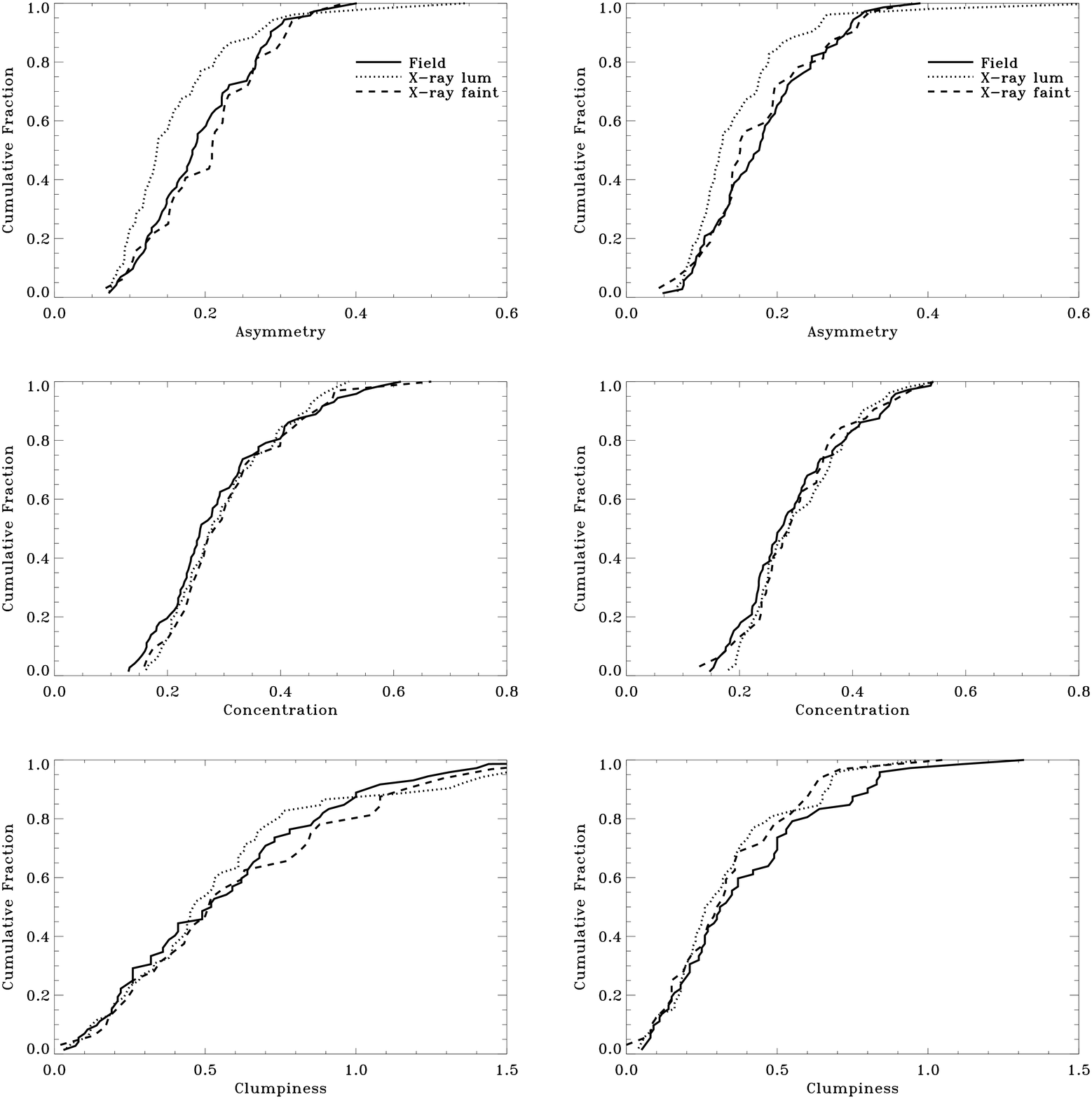}
\caption{Comparison of the individual CAS parameters in the blue (left) and
red (right) filters
for field (solid line), CL1604+4304 and CL1604+4321 (dashed line), and
CL0152-1357 and MS1054 (dotted line). There are marginally significant 
($2-2.5\sigma$) differences in the asymmetry distributions of the 
combined CL0152,MS1054 and 
CL1604+4304,+4321 samples (respectively, the X-ray luminous and X-ray 
faint clusters). There are also significant differences in the asymmetry 
distributions of the X-ray luminous cluster galaxies and the field galaxies,
which is related to the rest-frame color difference. See text for a discussion.
  \label{fig:sacblue}}
\end{center}
\end{figure*}

Depending on the physical processes affecting cluster galaxies, 
we might expect offsets from field galaxies 
in one or more of the CAS parameters. 
For example, in the low-redshift universe, there is 
evidence that spiral galaxies in
clusters are smoother than their field counterparts 
\citep{Gotoetal03,Mcintoshetal04} (S~$\downarrow$), but some
galaxies also have enhanced central star formation 
\citep{MW00,Mcintoshetal04,KK04},
which may lead to greater concentration (C~$\uparrow$). If 
galaxy-galaxy interactions are 
important in clusters, cluster galaxies will have greater asymmetries
(A~$\uparrow$).

In Figure~\ref{fig:sacblue} we show histograms
of the CAS parameters for galaxies in the field
(solid lines), the X-ray faint clusters (CL1604+4304 and CL1604+4321, 
dashed lines), and the X-ray luminous
clusters (CL0152 and MS1054; dotted lines). 
The distributions of C and S are indistinguishable between any of
the groupings (cluster/field, X-ray faint/luminous, 
X-ray faint/field, X-ray luminous/field) in the blue and red filters 
(rest-frame $U$ and $B$). In both blue and red filters there are 
marginally significant differences between the 
asymmetry distributions of the combined cluster and field samples, and 
between the X-ray luminous and X-ray faint clusters. 

In the blue filter, the late-type
galaxies in the field are more asymmetric than the late-type galaxies
in the X-ray luminous clusters (98\%), and the late-type galaxies in the 
X-ray faint clusters are more asymmetric than those in the X-ray 
luminous clusters (99\%). These trends are also marginally significant
in the red filter: 
X-ray luminous vs. field (99.2\%), X-ray luminous vs. X-ray faint (92\%).
There is no significant difference in the asymmetry distributions of
the field and the X-ray faint clusters.

We tested whether the difference in asymmetry between the two cluster samples
could be due to the difference
in filters; F606W is used for CL1604+4304,+4321, and MS1054, while F625W
is used for CL0152. A comparison of MS1054 and CL0152, both at $z=0.84$, 
with a K-S test shows that we cannot rule out that they are drawn
from the same parent population. Therefore, the asymmetry difference 
cannot be attributed to a difference in filter characteristics.

To recap, the late-type galaxies in the X-ray faint clusters tend to be 
more asymmetric than their counterparts in the X-ray luminous clusters.
If these trends are confirmed after the cluster and field sample
sizes are increased, then one possibility for the higher asymmetry values 
is a greater importance of galaxy-galaxy interactions in
these low-mass clusters and in lower density environments, presumably due
to the lower relative velocities. The X-ray luminous clusters also have 
asymmetry distributions that are significantly different than the field,
however, in the next section we show that this is related to the color
difference between the two samples.

\subsubsection{Asymmetry and Rest-frame Color}

\begin{figure*}
\begin{center}
\includegraphics[width=18cm]{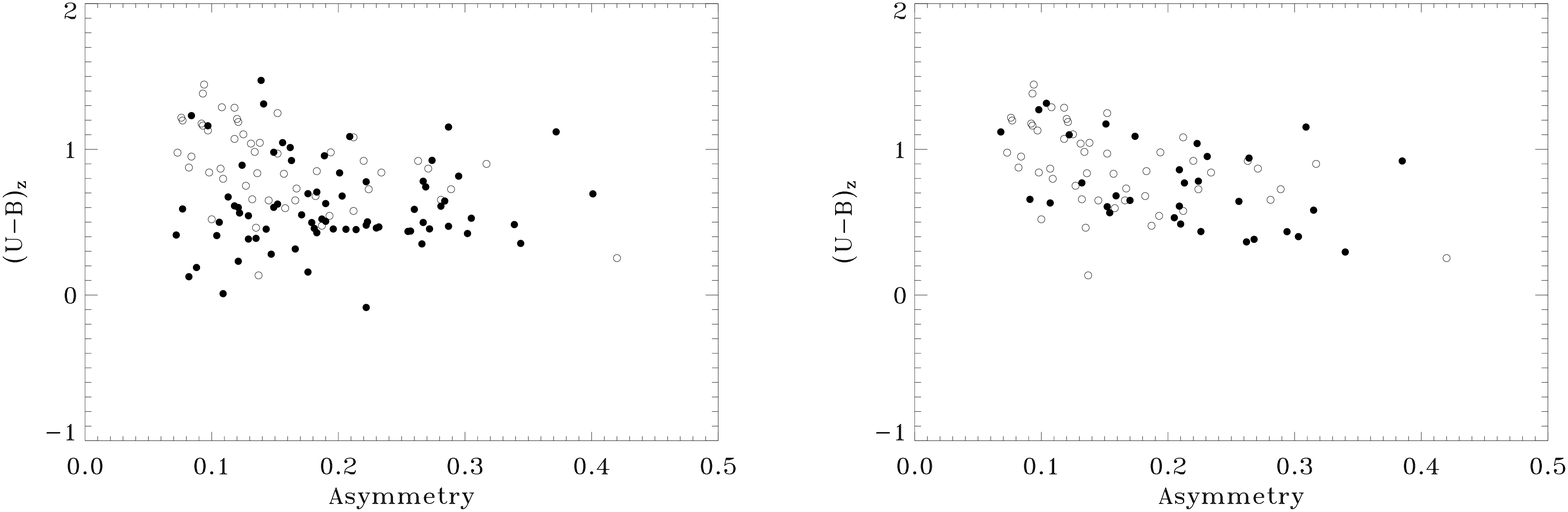}
\caption{Left panel: Comparison of the field (solid dots) and the 
X-ray luminous cluster galaxies 
(open dots) asymmetry and rest-frame $U-B$ color. Right panel: Comparison
of the X-ray faint (solid dots) and X-ray luminous cluster galaxies (open dots). When galaxies bluer than
$U-B=1$ are compared, no significant difference in asymmetry between the
two samples is found. 
  \label{fig:colorasymm}}
\end{center}
\end{figure*} 

We find a significant difference in asymmetry between the field
and X-ray luminous cluster galaxy samples. However, in \S~3.1 we 
showed that there is a significant rest-frame color difference
between the cluster and field samples, and in this section we show that this
is related to the offset in asymmetry between the X-ray luminous clusters and
the field. In Figure~\ref{fig:colorasymm} we show rest-frame $U-B$ color vs. 
asymmetry measured in the blue filter for the field (solid dots) and 
cluster (open dots) galaxies. These two parameters are not independent
for the cluster sample. The reddest cluster members have low asymmetries,
as might be expected. If we now compare only galaxies bluer than $U-B=1$,
the asymmetry distributions of the field and X-ray luminous cluster samples
are indistinguishable. But the asymmetry distributions of the 
X-ray faint and X-ray luminous clusters are still significantly different 
(98.7\%).

\subsubsection{Asymmetry and Local Galaxy Density}

\begin{figure}
\begin{center}
\includegraphics[width=9cm]{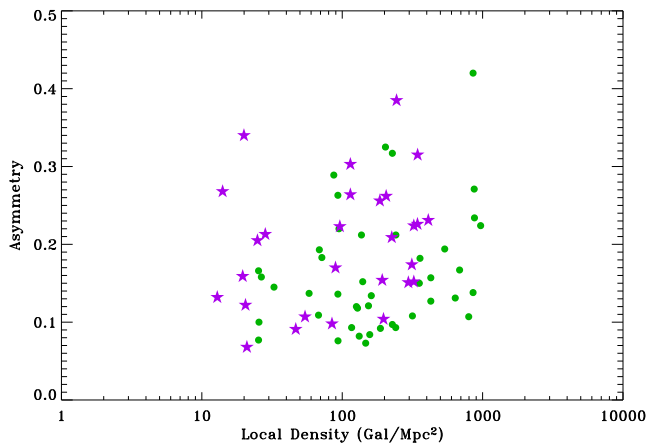}
\caption{Asymmetry values vs. local galaxy density in units of galaxies per
Mpc$^{-2}$. Dots are CL0152 and MS1054 cluster galaxies, and stars
are CL1604+4304 and +4321 cluster galaxies.
Local galaxy density measurements are from \citet{Postmanetal05}.
The densities are calculated with a statistical background 
subtraction.
  \label{asymmden}}
\end{center}
\end{figure}

Here we attempt to
determine if the asymmetry difference between the two cluster samples
could be due to variations in
local galaxy density, as measured by \citet{Postmanetal05}. In 
Figure~\ref{asymmden} we plot rest-frame $U$ asymmetry versus local galaxy 
density for 
the X-ray faint (stars) and X-ray luminous (filled circles) clusters.
The local galaxy densities were measured using a statistical
background subtraction, which is described in \citet{Postmanetal05}.
There is no statistically significant difference between the local
galaxy densities of the two samples.

\subsection{Physical Sizes}

In this section we compare the sizes of cluster and field late-type galaxies.
We might expect that disk galaxies in  
clusters will be smaller due to stripping of stars from
galaxy harassment or galaxy-galaxy interactions, a direct effect,
or indirectly from the stripping of disk gas, preventing future star formation.
Indeed, there is some observational
evidence that galaxies in the Coma cluster have smaller disks than
field galaxies \citep{Gutierrezetal04}.

We compared galaxy sizes in two ways. First, we compared half-light
radii from Sextractor, defined as the radius of a circular aperture 
which encloses
50\% of the light. The mean and median half-light radii for all
samples in {\it both} blue and red filters is approximately 3~kpc
(no correction for the PSF was made), with no significant differences
between the samples. 

\begin{figure*}
\begin{center}
\includegraphics[width=18cm]{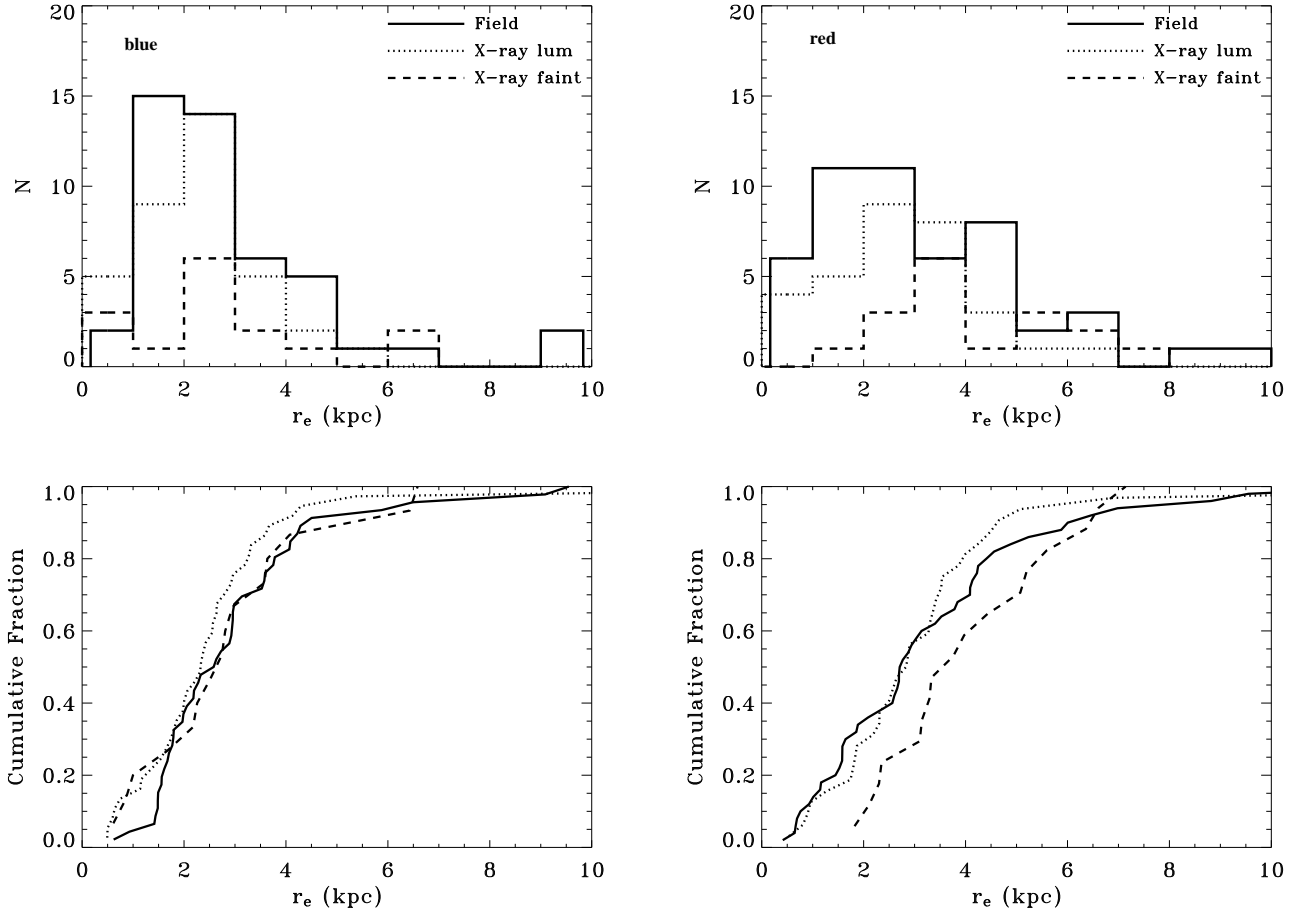}
\caption{Comparison of disk scale lengths in the blue (left panel) and
red (right panel) filters for spiral galaxies in the
field (solid line), CL1604+4304 and CL1604+4321 (dashed line), and
CL0152-1357 and MS1054 (dotted line). There are no statistically 
significant differences between any of the samples, and also note that
our sample size is small.
  \label{fig:disksizes}}
\end{center}
\end{figure*}

We also fit PSF-convolved 2D bulge+disk models to the spiral (but 
not irregular) galaxies using the GALFIT routine \citep{Petal02}.
However, as one can see from the color
cutouts in the Appendix, even the spiral galaxies have significant 
substructure. Many galaxies were not successfully fit with disk or 
bulge+disk profiles due to bright HII regions and asymmetric structure. 
This is because we are observing
in the rest-frame U and B, where star formation is most apparent.
For the galaxies which were successfully fit with disk or bulge+disk
models, we compare the disk scale lengths in Figure~\ref{fig:disksizes}.

In the blue filter, the results of K-S test comparison are that all
samples are consistent with being drawn from the same parent population.
In the red filter, a K-S test indicates that the probability
that the X-ray faint and field disk scale lengths are not drawn from the
same parent population is 92\%. We do not consider this to be significant,
and thus conclude that we find no
evidence for any significant differences in disk sizes between  
the clusters and the field, in either the rest-frame U or B. 
In summary, the sizes of field and cluster galaxies as measured by
half-light radii or disk scale length are indistinguishable. 

\subsection{Color Gradients}

There is evidence at $z<0.1$ that the color gradients of 
star-forming cluster and field galaxies differ, with field galaxies 
having red inward color gradients and cluster galaxies
having blue inward color gradients \citep{Mcintoshetal04}.
%the \citet{Mcintoshetal04} study selected galaxies based on 
%overall color in three local Abell clusters, A85, A496, and A754
This suggests that the cluster galaxies have larger nuclear star
formation rates relative to their outer disks, and is of interest 
in the discussion of the dominant
environmental processes that affect gas-rich cluster galaxies, as well as
the growth of bulges and central black holes.
Here we test whether the late-type cluster population at $z\sim0.9$ shows 
evidence for possessing the
same pattern of relatively enhanced nuclear star formation rate by examining
their color gradients. The difference in half-light radii at 
blue and red wavelengths gives a rough measure of the color gradient. 
We define the color gradient estimate (CGE),
CGE$=10\times$log$[\frac{r_{eff}^{red}}{r_{eff}^{blue}}]$
similar to \citet{Mcintoshetal04}. 

\begin{figure*}
\begin{center}
\includegraphics[width=18cm]{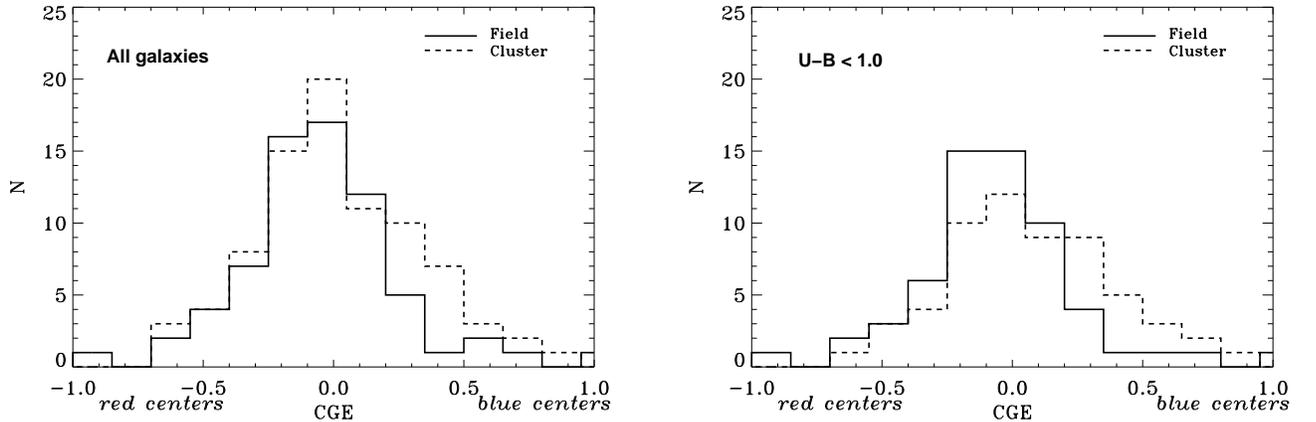}
\caption{Comparison of the field (solid line) and cluster (dashed line) 
color gradients. In the left panel, we compare the complete samples.
In the right panel, we only include those galaxies bluer than $U-B=1$, 
effectively removing the red cluster sequence galaxies. There is no
statistically significant difference between the complete
cluster and field samples, and a marginally significant difference in the
blue samples (97\%). The blue cluster galaxies tend to have blue inward 
color gradients. 
  \label{fig:cge}}
\end{center}
\end{figure*}

In the left panel of Figure~\ref{fig:cge} we show the CGE distributions 
for the cluster
(dashed line) and field galaxies (solid line). Interestingly, the field
galaxies tend to have red inward color gradients, 
similar to field galaxies at low redshifts \citep{deJong96,ME02},
most of which are late-type. It should be stressed that this rough 
estimate of color 
gradient measures color relatively within a given galaxy. For example,
if all cluster galaxies had positive values of CGE (blue inward) and 
all field galaxies had negative values (red inward), this would not 
necessarily mean that cluster galaxies have bluer centers than 
field galaxies, only that cluster galaxies have blue centers relative 
to the colors of their outer disks. 

From Figure~\ref{fig:cge} it appears that more cluster galaxies have 
blue inward color gradients, but the difference between the overall field and 
cluster populations is not statistically significant. A K-S test 
indicates that the two distributions are not drawn from the
same parent population at the 82\% level, which we do not consider to
be significant. The Wilcoxon rank-sum test, also referred to as 
the Mann-Whitney U-test, which is more sensitive to differences in the 
mean of distributions, finds a difference of approximately the same 
significance; the populations have a 85\% probability of not
having the same mean of distribution, which we also do not consider 
significant. If we choose our sample to include only those
galaxies with rest-frame $U-B < 1.0$, effectively removing the 
red cluster sequence members, then the difference in color gradient 
is much more significant, as shown in the right panel of
Figure~\ref{fig:cge}. Here a K-S test shows a significant (97\%)
difference between the distributions of CGE; more of the blue
cluster galaxies have blue inward color gradients. However, this
cluster sample is still significantly redder than the field sample.

Any difference in color gradient should be due to a difference
in the pattern of star formation. 
Relatively blue centers in the cluster galaxies could indicate 
enhanced nuclear 
star formation. The evidence for the existence of enhanced nuclear
star formation in Virgo spirals \citep{KK04}, late-type galaxies in 
Abell clusters \citep{MW00}, 
indicates that this is a possible interpretation of our results. 

We note that color gradient differences between cluster and 
field {\it early-type}, or 
spherodial, galaxies have also been found, but in an opposite sense.
Some fraction ($\sim 30$\%) of field 
early-type galaxies have blue inward color gradients that indicate 
a later or more extended formation epoch, 
whereas cluster early-type have the red inward color gradients expected
from metallicity gradients \citep{Menanteauetal01,Menanteauetal04}. 

In summary, compared to the field late-type galaxies, we find that 
more cluster 
late-types have blue centers relative to the color of their outer disks.
These relatively blue centers could indicate 
enhanced nuclear star formation rates, perhaps from gas driven in to the
galaxy centers from tidal forces. They could also be the result of 
truncated gas disks like those seen in Virgo spirals \citep{KK04}.

\section{Discussion and Conclusions}

We find a difference in the rest-frame $U$ and $B$ 
asymmetry distributions of the spiral 
and irregular galaxies in the X-ray luminous (CL0152 and MS1054)
and the X-ray faint (CL1604+4304 and CL1604+4321) clusters. The 
significance of these differences is marginal.
However, a visual examination of the galaxies
in these four clusters seems to confirm that at least the CL1604 system
includes more peculiar systems (such as ``comet-like'' shapes) than 
CL0152 and MS1054.

An increase in the sample
size of clusters is needed to confirm or refute this result, but if
confirmed, it would provide evidence for an increased importance of 
interactions in low-mass clusters and the field 
relative to high-mass clusters at
intermediate redshifts. Qualitatively, this may be plausible, 
as the relative velocities
should be smaller in less massive clusters ($\sigma_{v} \propto M^{1/2}$).
However, the magnitude of the difference in velocity dispersion between
these clusters is small, and complicated by substructure.
For example, the total velocity dispersion for CL0152 is 
$\sim 1600$~km~s$^{-1}$ \citep{Demarcoetal05}, but can be decomposed 
into three subclumps
with velocity dispersions ranging from $300-900$~km~s$^{-1}$ 
\citep{Girardietal05}.
The velocity dispersion for MS1054 is 
$\sim 1100-1200$~km~s$^{-1}$ \citep{Tranetal99}, 
while for CL1604+4304 and CL1604+4321  
the velocity dispersions are $\sim 960$ and $\sim 650$~km~s$^{-1}$ 
\citep{GL04}, respectively. Since CL1604+4321 dominates the X-ray 
faint cluster sample, the difference
in velocity dispersion between the two composite clusters (X-ray faint 
and X-ray luminous) is about a factor of 2. 

Most significantly, the color distributions of the cluster and 
field spiral/irregular galaxies differ, with the cluster sample 
being significantly redder. 
At the same time, we tentatively find that more cluster galaxies 
have blue inward color gradients,
possibly indicating enhanced central star formation relative
to the outer disk.

To interpret these results, it is useful to consider observations of
cluster spirals in the local universe, where we have 
the best chance of studying environmental processes in detail. 
Using H$\alpha$ imaging of 
Virgo cluster spirals, \citet{KK04} showed that the
reduction in overall SFR for spiral galaxies in local
clusters is due to a truncation of the gas disk. Star formation
rates in the centers of spiral galaxies in the cluster and field are similar,
but the lack of gas at outer radii in cluster spirals means that the
overall SFRs are suppressed. The physical mechanism identified as
responsible for
this gas disk truncation is ram pressure stripping, which could be 
aided by tidal loosening of the outer gas \citep{KK04}. 
Also, color gradients in low-redshift cluster and field galaxies with overall
blue colors differ, with field galaxies 
having red inward color gradients and cluster galaxies
having blue inward color gradients \citep{Mcintoshetal04}. This may be
circumstantial evidence that these blue cluster galaxies have relatively 
enhanced nuclear SFRs similar to Virgo spirals. 

There is other evidence that gas is driven to the centers of cluster spirals 
and causes circumnuclear starbursts. The study of low-redshift Abell clusters
by \citet{MW00} found that enhanced
nuclear star formation, as traced by H$\alpha$ emission, was correlated
with either a bar, or disturbed galaxy morphology, which they
concluded was evidence for ongoing tidal interactions. However, they found
an increase in galaxies classified as peculiar with increasing local 
galaxy density, something which is not consistent with the overall
morphology-density relation.

\citet{Tranetal05} studied the MS~2053 cluster system at $z=0.587$
and concluded that the spiral/merger galaxy population in MS~2053-B is 
indistinguishable from the field in terms of colors, luminosities, 
sizes, and [OII] $\lambda 3727$ EW. They conclude that this is an 
infalling field population.
However, the colors of the spiral/merger population in the 
more massive MS~2053-A are different than those in MS~2053-B and the field. 
They are redder on average, similar to what we find here. 
%Information on star formation rates from [OII] emission are readily
%available for CL1604+4304 and CL1604+4321 
%\citep{Lubinetal98,Postmanetal98,Postmanetal01}. 
%The median [OII] EW and SFR are not significantly different between the
%cluster and field late-type samples. However, this could be
%due to the small sample size. 

We interpret these findings as evidence that although some blue cluster
galaxies may be identified as infalling field galaxies, {\it as an 
ensemble, the late-type cluster members at intermediate redshifts are
not a pristine infalling field population.} There are significant
differences between cluster and field late-type galaxies
even at these redshifts.
Most of the cluster late-type galaxies are more than two
standard deviations away from the observed red cluster sequence, meaning that
these are Butcher-Oemler galaxies. Even given such blue colors, they
are still redder than field galaxies.

This may be further evidence for the separate evolution of color
and morphology (e.g. Goto et al. 2003, McIntosh et~al. 2004) in 
high density environments. The evolution in color occurs faster than 
the evolution in morphology.
This is puzzling, because the only mechanisms which would affect a galaxy's
color on a longer timescale than its morphology will also alter 
the {\it gas content}.
The only mechanisms which could alter the gas content and not the morphology 
(at least directly) are ram-pressure stripping and starvation. But 
these are not expected to be important because the morphology-density and
color-density relations are in place in low density regions where
ICM or intra-group medium pressure is negligible. If the ICM is unimportant,
but color differences occur in lower density regions than 
morphology differences, does this mean that nature is more important 
than nurture?

Our comparison of cluster and field late-type galaxies at intermediate redshift
shows that while the cluster galaxies are similar in most physical
parameters, they are signficantly redder. The color transformation is perhaps 
occuring at these or higher redshifts in lower density groups, with 
clusters accreting such groups with ``pre-processed'' galaxies
along filaments (e.g. Kodama et~al. 2001).

\smallskip
Our main conclusions are as follows. 

\begin{itemize}

\item{The late-type cluster population is redder than the late-type field
population. This is our most significant result. The mean rest-frame $U-B$
color difference is $0.34$ magnitudes between the combined cluster (X-ray 
luminous and X-ray faint) and field
samples. Although individual galaxies may be infalling from the field,
as an ensemble, they cannot be identified as a pristine infalling population.}

\item{At both rest-frame $U$ and $B$, we find a marginally
significant difference (98\%, 99\%) in the asymmetry distributions
of the late-type galaxies in the X-ray luminous clusters and the 
field. Late-type galaxies in 
X-ray luminous clusters have lower asymmetries than those in the field.
However, this difference can be completely attributed to the difference
in rest-frame color. Considering only those galaxies bluer than $U-B=1$, 
which removes the red cluster sequence galaxies, the asymmetry distributions
are indistinguishable.}

\item{At both rest-frame $U$ and $B$, we find a marginally
significant difference (99\%, 92\%) in the asymmetry distributions
of the late-type galaxies in the X-ray luminous clusters and the X-ray faint 
clusters. Galaxies in the X-ray
faint clusters have larger asymmetry values than those in X-ray luminous
clusters. This asymmetry difference persists when the red cluster sequence 
galaxies are removed and only the blue galaxies are considered (99\%).
If confirmed when the cluster and field sample sizes
are increased, this could point to a greater importance of interactions
in lower density regions at these redshifts. }

\item{We find no significant differences between any of the samples 
in concentration or clumpiness at rest-frame $U$ and $B$.}

\item{Physical sizes of field and cluster late-type galaxies are
similar. We find no significant difference in galaxy size as measured 
by half-light radii and disk scale lengths.}

\item{Considering only the blue ($U-B < 1$) cluster and field galaxies, we find
a marginally significant difference in the distributions of color gradients
for the cluster and field late-type populations. 
We find that more blue cluster galaxies have bluer inward gradients, possibly
indicating enhanced nuclear star formation as is seen in low redshift 
clusters \citep{KK04,Mcintoshetal04}.}

\end{itemize}

\smallskip

While similar in structure, physical size, and luminosity, the infalling
late-type galaxies in the outer cluster regions are redder and appear to 
have had a different star formation history than their late-type field
counterparts. This suggests that their assembly has already been influenced
by the somewhat denser environment in which they evolved.

\acknowledgements

We thank the referee for a careful reading and suggestions which improved 
the clarity of the manuscript.
We also thank T. Puzia for reading and commenting on an early version 
of this paper. 
ACS was developed under NASA contract NAS 5-32865, and this research 
has been supported by NASA grant NAG5-7697 and 
by an equipment grant from  Sun Microsystems, Inc.  
The {Space Telescope Science
Institute} is operated by AURA Inc., under NASA contract NAS5-26555.
We are grateful to K.~Anderson, J.~McCann, S.~Busching, A.~Framarini, 
S.~Barkhouser,
and T.~Allen for their invaluable contributions to the ACS project at JHU. 
We thank W.~J. McCann for the use of the FITSCUT routine for our color images.

\begin{deluxetable}{ccccccccccc}
\tablecolumns{11}
\tablewidth{0pc}
\tablecaption{CL1604+4304
\label{tab:4304cl}}
\tabletypesize{\scriptsize}
\tablehead{
\colhead{ACS ID} & \colhead{C 606} & \colhead{C 814} & \colhead{A 606}  & \colhead {A 814} & \colhead{S 606} &  \colhead{S 814} & \colhead{$r_{e}$ 606} & \colhead{$r_{e}$ 814} & \colhead{$v_{606}-i_{814}$} & \colhead{$i_{814}$}} 
\startdata
2933  & 0.251 & 0.239 & 0.210 & 0.156 & 0.119 & 0.099 & ---   & ---   &  0.73$\pm0.05$ & 24.29 $\pm0.04$\\
2531  & 0.466 & 0.474 & 0.209 & 0.151 & 0.032 & 0.012 & ***   & 5.9   &  1.25$\pm0.02$ & 22.37 $\pm0.01$\\
2930  & 0.219 & 0.256 & 0.309 & 0.309 & 0.128 & 0.049 & ---   & ---   &  1.59$\pm0.02$ & 21.84 $\pm0.01$\\
1495  & 0.326 & 0.337 & 0.151 & 0.149 & 0.091 & 0.046 & 16.9  & 18.3  &  1.63$\pm0.01$ & 20.85 $\pm0.00$\\
1627  & 0.401 & 0.349 & 0.340 & 0.304 & 0.087 & 0.073 & ---   & ---   &  0.54$\pm0.02$ & 22.53 $\pm0.02$\\
1448  & 0.283 & 0.260 & 0.268 & 0.197 & 0.051 & 0.041 & 9.2   & 11.4  &  0.64$\pm0.01$ & 22.24 $\pm0.01$\\
2121  & 0.496 & 0.438 & 0.154 & 0.140 & 0.030 & 0.021 & ---   & ---   &  0.86$\pm0.02$ & 22.68 $\pm0.01$\\
1135  & 0.315 & 0.336 & 0.385 & 0.379 & 0.028 & 0.011 & 16.5  & 17.0  &  1.33$\pm0.01$ & 21.99 $\pm0.01$\\
2701  & 0.667 & 0.543 & 0.209 & 0.183 & 0.003 & 0.002 & ---   & ---   &  0.94$\pm0.02$ & 22.99 $\pm0.01$
%9 galaxies
\enddata
\tablecomments{Asterisks indicate that 2D profile fitting with GALFIT was attempted, but did not converge. 
Effective radii are quoted in pixels (1~pixel=$0.05\arcsec$). Redshifts are given in the second column for field galaxies.}
\end{deluxetable}

\begin{deluxetable}{lccccccccccc}
\tablecolumns{12}
\tablewidth{0pc}
\tablecaption{CL1604+4304 field sample
\label{tab:4304f}}
\tabletypesize{\scriptsize}
\tablehead{
\colhead{ACS ID} & \colhead{redshift} & \colhead{C 606} & \colhead{C 814} & \colhead{A 606}  & \colhead {A 814} & \colhead{S 606} &  \colhead{S 814} & \colhead{$r_{e}$ 606} & \colhead{$r_{e}$ 814} & \colhead{$v_{606}-i_{814}$} & \colhead{$i_{814}$}}
\startdata
621  & 0.548 & 0.432 & 0.410 & 0.129 & 0.123 & 0.017 & 0.014 & 9.3  & 8.1  &  0.51$\pm0.01$ & 22.97 $\pm0.01$ \\
2166 & 0.609 & 0.361 & 0.364 & 0.196 & 0.149 & 0.048 & 0.040 & ---  & ---  &  0.62$\pm0.01$ & 22.26 $\pm0.01$\\
681  & 0.742 & 0.247 & 0.278 & 0.152 & 0.104 & 0.056 & 0.037 & 12.5 & 11.8 &  0.90$\pm0.02$ & 22.56 $\pm0.01$\\
2230 & 0.830 & 0.186 & 0.230 & 0.305 & 0.227 & 0.070 & 0.046 & ---  & ---  &  0.81$\pm0.01$ & 21.83 $\pm0.01$\\
1984 & 0.833 & 0.330 & 0.297 & 0.149 & 0.115 & 0.065 & 0.034 & 8.0  & 7.1  &  1.39$\pm0.02$ & 21.95 $\pm0.01$\\  
1656 & 0.937 & 0.309 & 0.315 & 0.082 & 0.074 & 0.017 & 0.018 & 6.6  & 6.9  &  0.32$\pm0.01$ & 22.42 $\pm0.01$\\
1389 & 0.973 & 0.219 & 0.237 & 0.122 & 0.098 & 0.070 & 0.052 & 9.0  & 12.4 &  0.80$\pm0.02$ & 22.53 $\pm0.01$\\
2645 & 0.974 & 0.292 & 0.338 & 0.129 & 0.091 & 0.033 & 0.022 & 7.4  & 7.7  &  0.83$\pm0.01$ & 22.47 $\pm0.01$\\
2539 & 0.985 & 0.289 & 0.364 & 0.156 & 0.098 & 0.082 & 0.033 & ---  & ---  &  1.42$\pm0.02$ & 21.74 $\pm0.01$\\
1068 & 1.068 & 0.326 & 0.286 & 0.084 & 0.103 & 0.088 & 0.037 & ---  & ---  &  1.61$\pm0.02$ & 22.23 $\pm0.01$\\
1374 & 1.085 & 0.613 & 0.538 & 0.176 & 0.244 & 0.004 & 0.005 & 4.4  & 7.2  &  0.24$\pm0.01$ & 22.78 $\pm0.01$\\ 
802  & 1.096 & 0.237 & 0.235 & 0.232 & 0.139 & 0.061 & 0.050 & 9.3  & 10.2 &  0.61$\pm0.02$ & 22.91 $\pm0.01$
%12 galaxies
\enddata
%\tablecomments{}
\end{deluxetable}

\begin{deluxetable}{lcccccccccc}
\tablecolumns{11}
\tablewidth{0pc}
\tablecaption{CL1604+4321
\label{tab:4321cl}}
\tabletypesize{\scriptsize}
\tablehead{
\colhead{ACS ID} & \colhead{C 606} & \colhead{C 814} & \colhead{A 606}  & \colhead {A 814} & \colhead{S 606} & \colhead{S 814} & \colhead{$r_{e}$ 606} & \colhead{$r_{e}$ 814} & \colhead{$v_{606}-i_{814}$} & \colhead{$i_{814}$}}
\startdata
1547  & 0.332 & 0.258 & 0.303 & 0.322 & 0.065 & 0.058 & ---  & ---   &  0.63$\pm0.02$ & 23.45 $\pm0.02$ \\
990   & 0.175 & 0.180 & 0.224 & 0.218 & 0.117 & 0.071 & ---  & ---   &  1.13$\pm0.02$ & 22.69 $\pm0.02$ \\
1849  & 0.165 & 0.163 & 0.294 & 0.226 & 0.120 & 0.096 & 5.8  & 13.3  &  0.72$\pm0.03$ & 23.69 $\pm0.03$ \\
880   & 0.338 & 0.347 & 0.104 & 0.101 & 0.096 & 0.035 & 2.2  & 13.0  &  1.84$\pm0.02$ & 21.95 $\pm0.01$ \\
883   & 0.267 & 0.309 & 0.231 & 0.195 & 0.114 & 0.061 & 2.6  & 8.0   &  1.37$\pm0.02$ & 22.21 $\pm0.01$ \\
1459  & 0.399 & 0.423 & 0.213 & 0.141 & 0.067 & 0.028 & 7.2  & 14.4  &  1.12$\pm0.01$ & 21.28 $\pm0.01$ \\
1640  & 0.210 & 0.293 & 0.174 & 0.140 & 0.136 & 0.060 & 1.6  &  ***  &  1.50$\pm0.01$ & 21.47 $\pm0.01$ \\
3443  & 0.272 & 0.287 & 0.132 & 0.093 & 0.033 & 0.021 & 4.3  & 4.7   &  1.07$\pm0.02$ & 23.05 $\pm0.01$ \\
2238  & 0.353 & 0.363 & 0.122 & 0.111 & 0.056 & 0.022 & 7.6  & 9.6   &  1.52$\pm0.02$ & 22.16 $\pm0.01$ \\
1892  & 0.491 & 0.502 & 0.098 & 0.043 & 0.030 & 0.007 & ***  & 6.0   &  1.75$\pm0.02$ & 21.87 $\pm0.01$ \\
1034  & 0.229 & 0.247 & 0.152 & 0.135 & 0.068 & 0.048 & 7.0  & ***   &  0.85$\pm0.02$ & 23.27 $\pm0.02$ \\
2195  & 0.305 & 0.275 & 0.226 & 0.192 & 0.028 & 0.025 & 5.6  & ***   &  0.70$\pm0.02$ & 22.97 $\pm0.01$ \\
2272  & 0.300 & 0.239 & 0.264 & 0.194 & 0.131 & 0.072 & ***  & 16.3  &  1.31$\pm0.02$ & 22.09 $\pm0.01$ \\
1341  & 0.244 & 0.196 & 0.091 & 0.079 & 0.036 & 0.028 & 6.5  & 5.4   &  0.98$\pm0.02$ & 23.35 $\pm0.02$ \\
2865  & 0.239 & 0.277 & 0.205 & 0.151 & 0.089 & 0.061 & 10.5 & 8.4   &  0.82$\pm0.02$ & 22.81 $\pm0.02$ \\
1930  & 0.200 & 0.216 & 0.159 & 0.112 & 0.053 & 0.037 & 9.3  & 8.5   &  1.01$\pm0.01$ & 21.96 $\pm0.01$ \\
1142  & 0.440 & 0.355 & 0.107 & 0.128 & 0.017 & 0.012 & ---  & ---   &  0.94$\pm0.02$ & 23.30 $\pm0.01$ \\
2470  & 0.235 & 0.307 & 0.068 & 0.064 & 0.119 & 0.042 & ***  & 10.2  &  1.50$\pm0.02$ & 22.24 $\pm0.01$ \\
2283  & 0.257 & 0.249 & 0.170 & 0.124 & 0.056 & 0.037 & ***  & 8.1   &  0.98$\pm0.02$ & 22.55 $\pm0.01$ \\
783   & 0.159 & 0.129 & 0.256 & 0.261 & 0.149 & 0.090 & ---  & ---   &  1.07$\pm0.04$ & 23.54 $\pm0.03$ \\
694   & 0.422 & 0.380 & 0.262 & 0.273 & 0.025 & 0.021 & ---  & ---   &  0.62$\pm0.01$ & 22.75 $\pm0.01$ \\
1160  & 0.296 & 0.293 & 0.315 & 0.262 & 0.047 & 0.036 & ---  & ---   &  0.89$\pm0.01$ & 22.08 $\pm0.01$ \\
2605  & 0.264 & 0.235 & 0.223 & 0.140 & 0.164 & 0.075 & ---  & ---   &  1.44$\pm0.02$ & 22.37 $\pm0.01$  
%23 galaxies
\enddata
%\tablecomments{}
\end{deluxetable}

\begin{deluxetable}{lccccccccccc}
\tablecolumns{12}
\tablewidth{0pc}
\tablecaption{CL1604+4321 field sample
\label{tab:4321f}}
\tabletypesize{\scriptsize}
\tablehead{
\colhead{ACS ID} & \colhead{redshift} & \colhead{C 606} & \colhead{C 814} & \colhead{A 606}  & \colhead {A 814} & \colhead{S 606} &  \colhead{S 814} & \colhead{$r_{e}$ 606} & \colhead{$r_{e}$ 814} & \colhead{$v_{606}-i_{814}$} & \colhead{$i_{814}$}}
\startdata
2338  & 0.618 & 0.407 & 0.414 & 0.121 & 0.093 & 0.015 & 0.011 & 8.9  & 8.1  & 0.77$\pm0.01$ & 21.56 $\pm0.01$\\
1071  & 0.618 & 0.459 & 0.417 & 0.211 & 0.145 & 0.009 & 0.010 & 5.9  & 4.7  & 0.44$\pm0.02$ & 23.38 $\pm0.02$\\
1596  & 0.694 & 0.548 & 0.467 & 0.295 & 0.279 & 0.076 & 0.054 & ---  & ---  & 1.00$\pm0.01$ & 21.34 $\pm0.01$\\
1655  & 0.695 & 0.288 & 0.309 & 0.281 & 0.235 & 0.051 & 0.037 & 8.3  & 7.4  & 0.84$\pm0.01$ & 22.40 $\pm0.01$\\
2143  & 0.696 & 0.293 & 0.267 & 0.267 & 0.214 & 0.056 & 0.047 & 25.8 & 25.1 & 0.71$\pm0.01$ & 21.59 $\pm0.01$\\
2140  & 0.715 & 0.248 & 0.253 & 0.162 & 0.117 & 0.056 & 0.032 & ***  & 12.4 & 1.32$\pm0.02$ & 22.29 $\pm0.01$\\
2220  & 0.726 & 0.350 & 0.335 & 0.255 & 0.206 & 0.032 & 0.027 & 6.4  & 4.6  & 0.64$\pm0.01$ & 22.46 $\pm0.01$\\
1572  & 0.726 & 0.219 & 0.266 & 0.072 & 0.048 & 0.034 & 0.025 & 5.6  & 5.2  & 0.64$\pm0.02$ & 23.32 $\pm0.01$\\  
1878  & 1.099 & 0.374 & 0.399 & 0.141 & 0.137 & 0.104 & 0.036 & 10.1 & 13.0 & 1.74$\pm0.02$ & 22.20 $\pm0.01$\\
1017  & 1.128 & 0.327 & 0.233 & 0.372 & 0.298 & 0.182 & 0.095 & 10.6 & 11.3 & 1.48$\pm0.02$ & 22.02 $\pm0.01$\\
966   & 1.174 & 0.239 & 0.231 & 0.222 & 0.181 & 0.040 & 0.036 & 8.8  & 10.1 & 0.85$\pm0.32$ & 27.48 $\pm0.21$\\
1438  & 1.182 & 0.252 & 0.175 & 0.267 & 0.301 & 0.150 & 0.104 & ---  & ---  & 0.91$\pm0.03$ & 23.44 $\pm0.03$
%12 galaxies
\enddata
%\tablecomments{}
\end{deluxetable}

\begin{deluxetable}{lcccccccccccc}
\tablecolumns{12}
\tablewidth{0pc}
\tablecaption{MS1054
\label{tab:1054cl}}
\tabletypesize{\scriptsize}
\tablehead{
\colhead{ACS ID} & \colhead{C 606} & \colhead{C 775} & \colhead{A 606}  & \colhead {A 775} & \colhead{S 606} &  \colhead{S 775} & \colhead{$r_{e}$ 606} & \colhead{$r_{e}$ 775} & \colhead{$v_{606}-i_{775}$} & \colhead{$i_{775}-z_{850}$} & \colhead{$i_{775}$}}
\startdata
392  & 0.293 & 0.329 & 0.182 & 0.150 & 0.081 & 0.034 & 8.6  & 8.8  &  0.83$\pm 0.02 $ & 0.17$\pm 0.02$ & 22.33$\pm 0.01$\\
456  & 0.521 & 0.411 & 0.107 & 0.087 & 0.065 & 0.029 & 1.6  & ***  &  0.96$\pm 0.03 $ & 0.34$\pm 0.03$ & 23.07$\pm 0.02$\\
835  & 0.196 & 0.220 & 0.289 & 0.256 & 0.144 & 0.090 & ---  & 18.1 &  0.67$\pm 0.03 $ & 0.10$\pm 0.03$ & 22.80$\pm 0.03$\\
947  & 0.186 & 0.194 & 0.098 & 0.068 & 0.075 & 0.035 & 6.2  & ***  &  1.51$\pm 0.02 $ & 0.54$\pm 0.01$ & 21.66$\pm 0.01$\\
921  & 0.351 & 0.364 & 0.108 & 0.124 & 0.103 & 0.032 & 6.7  & 6.9  &  0.93$\pm 0.04 $ & 0.20$\pm 0.03$ & 23.44$\pm 0.02$\\
959  & 0.388 & 0.335 & 0.194 & 0.175 & 0.160 & 0.056 & 1.3  & 1.4  &  1.21$\pm 0.03 $ & 0.43$\pm 0.02$ & 22.65$\pm 0.02$\\
1484 & 0.271 & 0.271 & 0.187 & 0.171 & 0.181 & 0.094 & ***  & 2.4  &  0.67$\pm 0.05 $ & 0.13$\pm 0.05$ & 23.71$\pm 0.04$\\
1948 & 0.242 & 0.199 & 0.220 & 0.179 & 0.135 & 0.079 & ***  & 54.0 &  1.08$\pm 0.01 $ & 0.27$\pm 0.01$ & 20.53$\pm 0.01$\\
2013 & 0.463 & 0.411 & 0.073 & 0.100 & 0.059 & 0.020 & 6.4  & 7.6  &  1.14$\pm 0.02 $ & 0.45$\pm 0.01$ & 22.62$\pm 0.01$\\
2210 & 0.432 & 0.465 & 0.093 & 0.110 & 0.021 & 0.006 & 4.7  & 6.5  &  1.39$\pm 0.02 $ & 0.48$\pm 0.01$ & 22.43$\pm 0.01$\\
2586 & 0.263 & 0.252 & 0.109 & 0.128 & 0.058 & 0.027 & 5.2  & ***  &  0.96$\pm 0.03 $ & 0.15$\pm 0.03$ & 23.00$\pm 0.02$\\
2813 & 0.297 & 0.341 & 0.076 & 0.068 & 0.136 & 0.035 & 1.7  & 9.2  &  1.47$\pm 0.02 $ & 0.61$\pm 0.01$ & 21.47$\pm 0.01$\\
2879 & 0.261 & 0.241 & 0.135 & 0.117 & 0.058 & 0.041 & 6.7  & ***  &  0.51$\pm 0.04 $ & 0.15$\pm 0.04$ & 23.87$\pm 0.03$\\
3070 & 0.229 & 0.250 & 0.167 & 0.140 & 0.115 & 0.062 & 5.4  & ***  &  1.12$\pm 0.03 $ & 0.39$\pm 0.02$ & 22.57$\pm 0.01$\\
2976 & 0.326 & 0.317 & 0.084 & 0.079 & 0.066 & 0.030 & 3.0  & ***  &  0.93$\pm 0.02 $ & 0.45$\pm 0.01$ & 21.51$\pm 0.01$\\
3075 & 0.355 & 0.382 & 0.077 & 0.096 & 0.039 & 0.016 & 9.4  & 13.3 &  1.44$\pm 0.01 $ & 0.43$\pm 0.00$ & 20.82$\pm 0.00$\\
3299 & 0.294 & 0.289 & 0.094 & 0.094 & 0.180 & 0.053 & ---  & ---  &  1.45$\pm 0.06 $ & 0.57$\pm 0.03$ & 23.30$\pm 0.03$\\
3322 & 0.216 & 0.266 & 0.100 & 0.117 & 0.040 & 0.019 & 4.3  & 4.7  &  0.67$\pm 0.03 $ & 0.09$\pm 0.03$ & 23.24$\pm 0.02$\\
3340 & 0.242 & 0.247 & 0.281 & 0.266 & 0.086 & 0.046 & ---  & ---  &  0.80$\pm 0.02 $ & 0.20$\pm 0.02$ & 22.49$\pm 0.02$\\
3403 & 0.346 & 0.359 & 0.193 & 0.201 & 0.040 & 0.021 & 4.5  & 4.9  &  0.68$\pm 0.02 $ & 0.24$\pm 0.02$ & 22.57$\pm 0.01$\\
3438 & 0.167 & 0.197 & 0.234 & 0.189 & 0.124 & 0.053 & 6.9  & 7.4  &  1.03$\pm 0.03 $ & 0.35$\pm 0.02$ & 22.62$\pm 0.02$\\
3727 & 0.203 & 0.207 & 0.137 & 0.154 & 0.112 & 0.083 & ---  & ---  &  0.34$\pm 0.03 $ & 0.06$\pm 0.04$ & 22.89$\pm 0.03$\\
3802 & 0.279 & 0.251 & 0.152 & 0.123 & 0.059 & 0.033 & 8.4  & 8.7  &  0.79$\pm 0.02 $ & 0.29$\pm 0.01$ & 22.28$\pm 0.01$\\
4101 & 0.230 & 0.240 & 0.263 & 0.221 & 0.175 & 0.074 & 5.8  & 6.7  &  1.02$\pm 0.03 $ & 0.37$\pm 0.01$ & 22.21$\pm 0.01$\\
4234 & 0.310 & 0.251 & 0.546 & 0.638 & 0.042 & 0.038 & ---  & ---  &  0.45$\pm 0.01 $ & 0.15$\pm 0.01$ & 21.07$\pm 0.00$\\
4777 & 0.279 & 0.263 & 0.152 & 0.177 & 0.061 & 0.034 & ---  & ---  &  1.17$\pm 0.01 $ & 0.32$\pm 0.01$ & 21.94$\pm 0.01$\\
5040 & 0.255 & 0.287 & 0.158 & 0.144 & 0.050 & 0.031 & 5.2  & ***  &  0.76$\pm 0.01 $ & 0.25$\pm 0.01$ & 21.51$\pm 0.01$\\
5152 & 0.190 & 0.202 & 0.082 & 0.127 & 0.118 & 0.055 & ---  & ---  &  1.15$\pm 0.03 $ & 0.31$\pm 0.02$ & 23.36$\pm 0.02$\\
5967 & 0.181 & 0.193 & 0.212 & 0.207 & 0.174 & 0.083 & ---  & 3.3  &  0.77$\pm 0.03 $ & 0.21$\pm 0.02$ & 22.37$\pm 0.02$\\
6567 & 0.224 & 0.225 & 0.166 & 0.159 & 0.088 & 0.045 & 6.1  & 9.0  &  0.78$\pm 0.02 $ & 0.27$\pm 0.02$ & 22.60$\pm 0.02$\\
6872 & 0.207 & 0.231 & 0.136 & 0.128 & 0.112 & 0.050 & 4.0  & 6.1  &  1.00$\pm 0.03 $ & 0.34$\pm 0.02$ & 22.43$\pm 0.02$\\
9133 & 0.381 & 0.380 & 0.120 & 0.111 & 0.108 & 0.034 & ***  & 6.0  &  1.34$\pm 0.03 $ & 0.46$\pm 0.01$ & 22.04$\pm 0.01$ 
%32 galaxies
\enddata
%\tablecomments{}
\end{deluxetable}

\begin{deluxetable}{lcccccccccccc}
\tablecolumns{13}
\tablewidth{0pc}
\tablecaption{MS1054 field sample
\label{tab:1054f}}
\tabletypesize{\scriptsize}
\tablehead{
\colhead{ACS ID} & \colhead{redshift} & \colhead{C 606} & \colhead{C 775} & \colhead{A 606}  & \colhead {A 775} & \colhead{S 606} &  \colhead{S 775} & \colhead{$r_{e}$ 606} & \colhead{$r_{e}$ 775} & \colhead{$v_{606}-i_{775}$} & \colhead{$i_{775}-z_{850}$} & \colhead{$i_{775}$}}
\startdata
1191  & 0.552 & 0.317 & 0.339 & 0.139 & 0.200 & 0.108 & 0.043 & 8.2  & 8.4   &  1.25$\pm 0.04 $ & 0.41$\pm 0.02$ & 22.84$\pm 0.02$\\
1542  & 0.553 & 0.252 & 0.303 & 0.149 & 0.159 & 0.091 & 0.035 & 3.7  & 16.2  & $0.560\pm 0.01$ & $0.12\pm 0.01$ & $21.60\pm 0.01$\\
556   & 0.559 & 0.279 & 0.280 & 0.176 & 0.205 & 0.056 & 0.038 & ---  & ---   &  0.75$\pm 0.01 $ & 0.21$\pm 0.01$ & 21.48$\pm 0.01$\\
1401  & 0.561 & 0.180 & 0.222 & 0.183 & 0.161 & 0.106 & 0.067 & ***  & 2.0   &  0.56$\pm 0.02 $ & 0.08$\pm 0.02$ & 21.66$\pm 0.01$\\
2478  & 0.573 & 0.361 & 0.310 & 0.077 & 0.076 & 0.046 & 0.027 & ---  & ---   &  0.73$\pm 0.02 $ & 0.20$\pm 0.02$ & 22.97$\pm 0.02$\\
531   & 0.580 & 0.230 & 0.257 & 0.344 & 0.316 & 0.109 & 0.087 & 20.0 & 19.9  &  0.48$\pm 0.02 $ &-0.03$\pm 0.02$ & 22.31$\pm 0.01$\\
5659  & 0.600 & 0.260 & 0.273 & 0.302 & 0.288 & 0.087 & 0.061 & ---  & ---   &  0.56$\pm 0.02 $ & 0.02$\pm 0.02$ & 21.99$\pm 0.01$\\
3573  & 0.617 & 0.223 & 0.223 & 0.109 & 0.081 & 0.067 & 0.053 & 7.9  & 9.0   &  0.21$\pm 0.02 $ & 0.12$\pm 0.03$ & 23.12$\pm 0.02$\\
3865  & 0.618 & 0.256 & 0.253 & 0.088 & 0.076 & 0.059 & 0.041 & 2.6  & 7.8   &  0.36$\pm 0.02 $ & 0.10$\pm 0.02$ & 22.23$\pm 0.01$\\
9272  & 0.666 & 0.470 & 0.447 & 0.106 & 0.087 & 0.007 & 0.004 & 3.8  & 3.0   &  0.67$\pm 0.01 $ & 0.15$\pm 0.01$ & 22.36$\pm 0.01$\\
3523  & 0.702 & 0.217 & 0.222 & 0.339 & 0.341 & 0.139 & 0.086 & ---  & ---   &  0.62$\pm 0.02 $ & 0.11$\pm 0.02$ & 22.20$\pm 0.02$\\
9214  & 0.702 & 0.234 & 0.242 & 0.183 & 0.176 & 0.079 & 0.046 & 9.4  & 1.8   &  0.88$\pm 0.01 $ & 0.26$\pm 0.01$ & 21.19$\pm 0.01$\\
8165  & 0.702 & 0.171 & 0.202 & 0.190 & 0.178 & 0.075 & 0.039 & ---  & ---   &  0.66$\pm 0.04 $ & 0.13$\pm 0.04$ & 23.80$\pm 0.03$\\ 
5464  & 0.715 & 0.323 & 0.370 & 0.124 & 0.130 & 0.049 & 0.025 & 5.6  & 4.0   &  1.07$\pm 0.01 $ & 0.45$\pm 0.01$ & 22.20$\pm 0.01$\\
6284  & 0.756 & 0.181 & 0.184 & 0.190 & 0.184 & 0.126 & 0.066 & 4.4  & 15.6  &  0.80$\pm 0.02 $ & 0.24$\pm 0.02$ & 22.04$\pm 0.01$\\
4324  & 0.759 & 0.462 & 0.464 & 0.209 & 0.183 & 0.021 & 0.014 & 3.4  & 3.5   &  1.30$\pm 0.02 $ & 0.41$\pm 0.01$ & 22.21$\pm 0.01$\\
1203  & 0.761 & 0.314 & 0.282 & 0.257 & 0.213 & 0.071 & 0.050 & 4.4  & 4.8   &  0.53$\pm 0.01 $ & 0.14$\pm 0.01$ & 21.55$\pm 0.01$\\
7248  & 0.761 & 0.234 & 0.320 & 0.163 & 0.137 & 0.147 & 0.057 & 4.5  & 1.9   &  1.13$\pm 0.01 $ & 0.43$\pm 0.01$ & 21.20$\pm 0.01$\\
7538  & 0.762 & 0.579 & 0.499 & 0.113 & 0.103 & 0.013 & 0.008 & 1.6  & 1.1   &  0.81$\pm 0.01 $ & 0.30$\pm 0.01$ & 22.57$\pm 0.01$\\
6320  & 0.762 & 0.168 & 0.187 & 0.401 & 0.390 & 0.113 & 0.067 & 4.1  & 34.9  &  0.94$\pm 0.02 $ & 0.29$\pm 0.01$ & 21.59$\pm 0.01$\\
3970  & 0.778 & 0.228 & 0.234 & 0.230 & 0.169 & 0.058 & 0.036 & 4.1  & ***   &  0.53$\pm 0.03 $ & 0.13$\pm 0.03$ & 23.51$\pm 0.02$\\
7987  & 0.870 & 0.199 & 0.212 & 0.266 & 0.292 & 0.069 & 0.050 & 7.7  & 3.1   &  0.47$\pm 0.01 $ & 0.18$\pm 0.01$ & 21.78$\pm 0.01$\\
4887  & 0.873 & 0.404 & 0.381 & 0.203 & 0.181 & 0.031 & 0.021 & 4.3  & 4.1   &  0.84$\pm 0.01 $ & 0.25$\pm 0.01$ & 22.25$\pm 0.01$\\ 
1560  & 0.896 & 0.155 & 0.198 & 0.269 & 0.264 & 0.246 & 0.116 & ---  & ---   &  0.95$\pm 0.02 $ & 0.34$\pm 0.01$ & 21.31$\pm 0.01$\\
305   & 0.897 & 0.252 & 0.265 & 0.149 & 0.131 & 0.078 & 0.039 & 5.7  & 7.0   &  0.75$\pm 0.03 $ & 0.22$\pm 0.02$ & 22.76$\pm 0.02$\\
7225  & 0.897 & 0.205 & 0.241 & 0.223 & 0.177 & 0.103 & 0.056 & 4.7  & 4.3   &  0.66$\pm 0.03 $ & 0.21$\pm 0.03$ & 23.25$\pm 0.02$\\
2643  & 0.897 & 0.133 & 0.147 & 0.260 & 0.244 & 0.189 & 0.105 & ---  & ---   &  0.71$\pm 0.03 $ & 0.33$\pm 0.03$ & 22.71$\pm 0.03$\\
6423  & 0.971 & 0.163 & 0.183 & 0.135 & 0.141 & 0.069 & 0.044 & 4.2  & ***   &  0.58$\pm 0.03 $ & 0.24$\pm 0.03$ & 23.49$\pm 0.03$\\
1668  & 0.978 & 0.131 & 0.176 & 0.143 & 0.189 & 0.214 & 0.101 & ---  & ---   &  0.78$\pm 0.06 $ & 0.35$\pm 0.05$ & 23.79$\pm 0.05$\\
1021  & 0.990 & 0.163 & 0.160 & 0.121 & 0.143 & 0.127 & 0.087 & 6.0  & 4.4   &  0.31$\pm 0.06 $ & 0.59$\pm 0.06$ & 24.35$\pm 0.07$\\
2991  & 1.060 & 0.223 & 0.157 & 0.287 & 0.310 & 0.373 & 0.130 & ---  & ---   &  1.19$\pm 0.09 $ & 0.76$\pm 0.04$ & 23.80$\pm 0.05$\\
5649  & 1.073 & 0.212 & 0.257 & 0.187 & 0.134 & 0.041 & 0.025 & ---  & ---   &  0.56$\pm 0.03 $ & 0.37$\pm 0.03$ & 23.67$\pm 0.03$\\
3881  & 1.075 & 0.501 & 0.476 & 0.272 & 0.297 & 0.029 & 0.025 & ---  & ---   &  0.51$\pm 0.01 $ & 0.24$\pm 0.01$ & 21.32$\pm 0.00$\\
3046  & 1.076 & 0.142 & 0.190 & 0.189 & 0.198 & 0.078 & 0.049 & ---  & ---   & $1.08\pm 0.05$ & $0.71\pm 0.03$ & $23.44\pm 0.04$\\
1289  & 1.140 & 0.241 & 0.258 & 0.166 & 0.197 & 0.055 & 0.040 & 5.3  & 10.1  &  0.33$\pm 0.03 $ & 0.41$\pm 0.03$ & 23.38$\pm 0.03$
%35 galaxies
\enddata
%\tablecomments{}
\end{deluxetable}

\begin{deluxetable}{lcccccccccccc}
\tablecolumns{12}
\tablewidth{0pc}
\tablecaption{CL0152-1357
\label{tab:0152cl}}
\tabletypesize{\scriptsize}
\tablehead{
\colhead{ACS ID} & \colhead{C 625} & \colhead{C 775} & \colhead{A 625}  & \colhead {A 775} & \colhead{S 625} &  \colhead{S 775} & \colhead{$r_{e}$ 625} & \colhead{$r_{e}$ 775}& \colhead{$r_{625}-i_{775}$} & \colhead{$i_{775}-z_{850}$} & \colhead{$i_{775}$} } 
\startdata
10063 & 0.402 & 0.392 & 0.134 & 0.109 & 0.070 & 0.039 & ---   & ---  &  0.95$\pm 0.02 $ & 0.44$\pm 0.02$ & 22.65$\pm 0.01$\\
11019 & 0.430 & 0.496 & 0.097 & 0.075 & 0.022 & 0.007 & ---   & ---  &  1.09$\pm 0.02 $ & 0.65$\pm 0.01$ & 22.68$\pm 0.01$\\
11613 & 0.271 & 0.287 & 0.118 & 0.107 & 0.062 & 0.030 & ---   & ---  &  1.04$\pm 0.03 $ & 0.48$\pm 0.02$ & 23.06$\pm 0.01$\\
1146  & 0.217 & 0.239 & 0.127 & 0.122 & 0.038 & 0.038 & 7.8   & 12.2 &  0.74$\pm 0.02 $ & 0.32$\pm 0.02$ & 22.58$\pm 0.02$\\ 
1564  & 0.392 & 0.361 & 0.183 & 0.189 & 0.056 & 0.037 & 7.6   & 4.6  &  0.78$\pm 0.02 $ & 0.33$\pm 0.02$ & 22.66$\pm 0.02$\\
1575  & 0.314 & 0.354 & 0.125 & 0.084 & 0.045 & 0.036 & 4.9   & ***  &  1.06$\pm 0.04 $ & 0.70$\pm 0.02$ & 22.88$\pm 0.02$\\
1652  & 0.247 & 0.263 & 0.157 & 0.119 & 0.063 & 0.044 & 8.7   & 8.6  &  0.80$\pm 0.01 $ & 0.28$\pm 0.01$ & 21.63$\pm 0.01$\\
1737  & 0.207 & 0.215 & 0.131 & 0.114 & 0.077 & 0.042 & 3.0   & 2.1  &  1.03$\pm 0.02 $ & 0.42$\pm 0.02$ & 22.79$\pm 0.01$\\
2016  & 0.450 & 0.412 & 0.420 & 0.404 & 0.052 & 0.050 & ---   & ---  &  0.26$\pm 0.01 $ & 0.12$\pm 0.02$ & 22.28$\pm 0.01$\\
2027  & 0.240 & 0.302 & 0.317 & 0.246 & 0.094 & 0.055 & 9.6   & 11.2 &  0.88$\pm 0.02 $ & 0.42$\pm 0.01$ & 22.13$\pm 0.01$\\
2235  & 0.453 & 0.457 & 0.118 & 0.106 & 0.056 & 0.023 & 14.1  & 4.8  &  1.24$\pm 0.02 $ & 0.60$\pm 0.01$ & 21.70$\pm 0.01$\\
3329  & 0.645 & 0.539 & 0.352 & 0.300 & 0.066 & 0.049 & ---   & ---  &  0.80$\pm 0.01 $ & 0.62$\pm 0.00$ & 20.62$\pm 0.00$\\
3390  & 0.317 & 0.349 & 0.121 & 0.103 & 0.126 & 0.052 & 1.3   & 2.5  &  1.20$\pm 0.02 $ & 0.70$\pm 0.01$ & 21.07$\pm 0.01$\\
3927  & 0.265 & 0.292 & 0.271 & 0.260 & 0.087 & 0.061 & 10.8  & 11.8 &  0.84$\pm 0.01 $ & 0.33$\pm 0.01$ & 21.57$\pm 0.01$\\
5410  & 0.478 & 0.439 & 0.138 & 0.138 & 0.048 & 0.024 & 6.1   & 9.3  &  1.02$\pm 0.01 $ & 0.48$\pm 0.01$ & 21.59$\pm 0.01$\\
5481  & 0.329 & 0.387 & 0.212 & 0.174 & 0.057 & 0.031 & 3.5   & 10.5 &  1.04$\pm 0.01 $ & 0.63$\pm 0.01$ & 22.02$\pm 0.01$\\
7017  & 0.393 & 0.420 & 0.092 & 0.077 & 0.049 & 0.027 & 2.0   & 7.5  &  1.14$\pm 0.01 $ & 0.57$\pm 0.01$ & 21.25$\pm 0.01$\\
717   & 0.317 & 0.327 & 0.093 & 0.098 & 0.101 & 0.043 & 7.3   & ***  &  1.26$\pm 0.03 $ & 0.60$\pm 0.02$ & 22.43$\pm 0.02$\\
8671  & 0.207 & 0.234 & 0.224 & 0.187 & 0.078 & 0.057 & 11.2  & 10.1 &  0.72$\pm 0.01 $ & 0.30$\pm 0.01$ & 21.41$\pm 0.01$\\
8708  & 0.162 & 0.180 & 0.132 & 0.082 & 0.070 & 0.043 & 6.9   & 5.7  &  0.57$\pm 0.03 $ & 0.26$\pm 0.03$ & 23.44$\pm 0.02$\\
9563  & 0.498 & 0.546 & 0.152 & 0.087 & 0.029 & 0.013 & ---   & ---  &  1.23$\pm 0.02 $ & 0.57$\pm 0.01$ & 22.08$\pm 0.01$
%20 galaxies
\enddata
%\tablecomments{}
\end{deluxetable}

\begin{deluxetable}{lcccccccccccc}
\tablecolumns{13}
\tablewidth{0pc}
\tablecaption{CL0152 field sample
\label{tab:0152f}}
\tabletypesize{\scriptsize}
\tablehead{
\colhead{ACS ID} & \colhead{redshift} & \colhead{C 606} & \colhead{C 775} & \colhead{A 606}  & \colhead {A 775} & \colhead{S 606} &  \colhead{S 775} & \colhead{$r_{e}$ 625} & \colhead{$r_{e}$ 775} & \colhead{$r_{625}-i_{775}$} & \colhead{$i_{775}-z_{850}$} & \colhead{$i_{775}$}}
\startdata
668   & 0.556 & 0.399 & 0.388 & 0.171 & 0.168 & 0.027 & 0.021 & ---   & ---  &  0.46$\pm 0.01 $ & 0.22$\pm 0.01$ & 21.35$\pm 0.01$\\
1051  & 0.670 & 0.406 & 0.449 & 0.222 & 0.211 & 0.079 & 0.054 & 5.0   & 5.2  &  0.75$\pm 0.01 $ & 0.35$\pm 0.01$ & 21.24$\pm 0.01$\\
2199  & 0.675 & 0.242 & 0.229 & 0.222 & 0.246 & 0.072 & 0.084 & 12.3  & 12.2 & -0.01$\pm 0.03 $ &-0.03$\pm 0.05$ & 24.29$\pm 0.04$\\
978   & 0.720 & 0.280 & 0.316 & 0.118 & 0.120 & 0.033 & 0.025 & 10.1  & 8.8  &  0.58$\pm 0.01 $ & 0.19$\pm 0.01$ & 21.91$\pm 0.01$\\
2399  & 0.739 & 0.474 & 0.469 & 0.104 & 0.088 & 0.029 & 0.025 & ---   & ---  &  0.37$\pm 0.00 $ & 0.28$\pm 0.01$ & 21.11$\pm 0.00$\\
2456  & 0.746 & 0.271 & 0.305 & 0.214 & 0.192 & 0.038 & 0.031 & 11.1  & 11.3 &  0.42$\pm 0.02 $ & 0.23$\pm 0.03$ & 23.70$\pm 0.02$\\
5222  & 0.788 & 0.494 & 0.541 & 0.097 & 0.092 & 0.029 & 0.009 & 2.3   & 2.5  &  1.16$\pm 0.01 $ & 0.58$\pm 0.01$ & 21.64$\pm 0.01$\\
8851  & 0.959 & 0.160 & 0.154 & 0.274 & 0.266 & 0.148 & 0.093 & ---   & ---  &  0.84$\pm 0.04 $ & 0.47$\pm 0.03$ & 23.12$\pm 0.03$\\
5804  & 0.960 & 0.278 & 0.298 & 0.201 & 0.163 & 0.066 & 0.043 & 12.6  & 9.7  &  0.79$\pm 0.01 $ & 0.43$\pm 0.01$ & 21.70$\pm 0.01$\\
805   & 0.993 & 0.534 & 0.456 & 0.206 & 0.219 & 0.030 & 0.024 & 1.7   & ***  &  0.48$\pm 0.01 $ & 0.41$\pm 0.01$ & 21.72$\pm 0.01$\\
8441  & 1.048 & 0.414 & 0.391 & 0.179 & 0.152 & 0.037 & 0.032 & 14.9  & 14.7 &  0.47$\pm 0.02 $ & 0.40$\pm 0.02$ & 22.71$\pm 0.01$\\
6851  & 1.101 & 0.257 & 0.234 & 0.287 & 0.246 & 0.077 & 0.066 & ---   & ---  &  0.42$\pm 0.01 $ & 0.42$\pm 0.01$ & 22.25$\pm 0.01$\\
6879  & 1.120 & 0.149 & 0.167 & 0.284 & 0.280 & 0.119 & 0.095 & 24.0  & 23.6 &  0.55$\pm 0.01 $ & 0.62$\pm 0.01$ & 21.94$\pm 0.01$
%12 galaxies
\enddata
%\tablecomments{}
\end{deluxetable}

\end{document}